\documentclass[fleqn,12pt,twoside]{elsart}

\usepackage{graphicx}
\usepackage{amssymb,amsmath}

\usepackage{citesort}

\newcommand{\be}{\begin{eqnarray}}
\newcommand{\ee}{\end{eqnarray}}

\newcommand{\pme}[2]{\ensuremath{\,\pm\,{}^{#1}_{#2}}}

\newcommand{\eqeqref}[1]{Eq.~\eqref{#1}}

\newcommand{\refref}[1]{Ref.~\cite{#1}}

\newcommand{\secref}[1]{Sec.~\ref{#1}}

\newcommand{\figref}[1]{Fig.~\ref{#1}}
\newcommand{\tabref}[1]{Table~\ref{#1}}

\newcommand{\beq}{\begin{equation}}
\newcommand{\eeq}{\end{equation}}
\newcommand{\bea}{\begin{eqnarray}}
\newcommand{\beas}{\begin{eqnarray*}}
\newcommand{\beau}[1]{\begin{equation} \label{#1} \begin{array}{rcl}}
\newcommand{\eea}{\end{eqnarray}}
\newcommand{\eeas}{\end{eqnarray*}}
\newcommand{\eeau}{\end{array} \end{equation}}
\newcommand{\bay}{\begin{array}}
\newcommand{\eay}{\end{array}}
\newcommand{\bals}{\begin{align*}}
\newcommand{\eals}{\end{align*}}

\newcommand{\lora}{{\longrightarrow}}

\newcommand{\ra}{{\rightarrow}}

\newcommand{\vev}[1]{\langle #1 \rangle}

\newcommand{\OO}{{\mathcal O}}
\newcommand{\PP}{{\mathcal P}}

\begin{document}

\begin{frontmatter}



  \title{ Atomic Mass Dependence of Hadron Production in
    Deep Inelastic Scattering on  Nuclei}

  \author[Columbia,ISU]{A.~Accardi}
  , \author[Heidelberg]{D.~Gr\"unewald\thanksref{Grunemail}}
  , \author[LNF]{V.~Muccifora} 
  and \author[Heidelberg]{H.J.~Pirner} 


  \address[Columbia]{Columbia Physics Department, 538 West 120th
    Street, New York, NY 10027, U.S.A.}
  \address[ISU]{Iowa State University, Dept. of Physics and
    Astrophysics, Ames, IA 50011, U.S.A.}
  \address[Heidelberg]{Institut f\"ur Theoretische Physik der
    Universit\"at Heidelberg Philosophenweg 19, D-69120 Heidelberg, Germany }
  \address[LNF]{INFN, Laboratori Nazionali di Frascati, I-00044
    Frascati, Italy }
  \thanks[Grunemail]{Corresponding author. E-mail address: daniel@tphys.uni-heidelberg.de}

  \begin{abstract} 
Hadron production in lepton-nucleus deep inelastic scattering is
studied in an absorption model. In the proposed model, 
the early stage of
hadronization in the nuclear medium is dominated by prehadron
formation and absorption, controlled by flavor-dependent formation
lengths and absorption cross sections. 
Computations for hadron
multiplicity ratios are presented and compared with the HERMES
experimental data for pions, kaons, protons and
antiprotons. The mass-number dependence of hadron attenuation is shown
to be sensitive to the underlying hadronization dynamics. 
Contrary to common expectations for absorption models, a leading term
proportional to $A^{2/3}$ is found. 
Deviations from the leading behavior arise at large mass-numbers and
large hadron fractional momenta.  

  \end{abstract} 

  
\end{frontmatter}


\section{Introduction}
\label{sec:introduction}

Recent HERMES results give  precise data on hadron production in 
deep inelastic scattering (DIS) of 27.6 GeV positrons  on D, He, N,
Ne, and Kr nuclei \cite{HERM01,HERM03,HERM04}.
The main observable is the multiplicity ratio, $R_M$, defined
as the ratio between the hadron multiplicity on 
nucleus and on deuterium.
$R_M$ has been  studied as a function of the hadron fractional momentum $z$,
of the virtual photon energy $\nu$, of its virtuality $Q^2$,  for different hadrons and for different nuclei. 
The use of nuclear targets allows to study the hadronization process 
near the interaction point 
of the photon and probe hadron formation only some  
fermi away.

There are many proposed models to describe how hadronization in
presence of a nucleus evolves in space and time.  
The model computations range from a color string 
breaking mechanism~\cite{BI80,BG87,BICH83,GY90,CZ92,AK02} with final
state interaction 
~\cite{FAL03,FALTER04}, to gluon bremsstrahlung for leading
hadrons~\cite{boris,KOP03} and pure energy loss
models~\cite{WW02,MWW04,arleo02}. 
Two classes of models of hadron formation compete with each other.
The first one is based on nuclear absorption, 
where the color of the struck quark is neutralized 
after a short time by the formation of a prehadron, the predecessor of the
final hadron, which interacts with surrounding nucleons and 
is absorbed on its way out of the nucleus.
The second model assumes  
energy loss from medium-induced radiation of the struck quark
as dominant process.
The first interpretation emphasizes the hadronic aspects of particle
production, 
the second one focuses on partonic degrees of freedom and 
postpones hadronization outside the nucleus. Neglecting the
production of secondary particles, both
mechanisms reduce the number of hadrons emerging from the nucleus.
The current belief is that the  dependence of  hadron attenuation
from the mass-number $A$ of the target nucleus
can differentiate the two processes: in the absorption model,
hadron attenuation is commonly believed to be  
proportional to the path length of the (pre)hadron in the nucleus
($\propto A^{1/3}$), whereas  in the energy loss model the 
attenuation is supposed to depend on the square of the distance the quark
traverses in the nucleus ($\propto A^{2/3}$).
A careful measurement of the $A$-dependence of the nuclear attenuation
would therefore allow  to discriminate between the two different processes.

In a previous paper~\cite{ACC02}, we have calculated  the nuclear
modifications of hadron  production in DIS in an absorption model. The hadron formation length
has been  computed analytically in the framework of the Lund string
model \cite{BG87,AND83} as a two-step process.
In the first step 
a quark-antiquark pair  from the break-up of the color string
forms a prehadron. In the second step
the final state hadron is created. Both the prehadron and the hadron,
if they form inside the nucleus,
interact with target nucleons  and may be absorbed.
Setting the prehadron nucleon cross section equal to the hadron nucleon
cross section in \refref{ACC02} we had to  increase the formation
length by using an effective string tension $\kappa$=0.4 GeV/fm, much
smaller than the expected $\kappa \approx 1$ GeV/fm (see \refref{MAR03}
for example).

In this paper we  correct the model  
by combining  a realistic formation length based on the expected
string constant with smaller prehadronic cross sections. 
We pay special attention to the 
flavor dependence of the  formation length,
which is naturally induced by the Lund model,
and calculate the multiplicity ratio $R_M^h$ for different
hadron species $h$ as functions of kinematic variables $z$ and $\nu$. Finally, we study the $A$-dependence of the nuclear
attenuation $1-R_M^h$, and challenge common expectations by showing,
analytically and numerically, a $A^{2/3}$ dependence for the
absorption model.  

The paper is organized as follows. 
In section~\ref{sec:DevModel}, the new developments of the model are
described. Section~\ref{sec:A-dependence} is 
devoted to the analytic calculation of the $A$-dependence of hadron
attenuation.
In Section~\ref{sec:results}, numerical
computations of $R_M^h$ and of the $A$-dependence of $1-R_M$ 
are presented and compared with experimental data. 
Section~\ref{sec:conclusions} presents a summary and our
conclusions.


\section{Development of the model}
\label{sec:DevModel}
In this section we report the  main developments of the model, which
has been extensively described in \refref{ACC02}.
The experimental multiplicity ratio $R_M^h$ between nucleus and deuterium 
is a double ratio and can be measured as a function of $z$, $\nu$ or
$Q^2$.
As a function of $z$ it is given by:
\begin{align}
  R_M^h(z) = \frac{1}{N_A^{DIS}}\frac{dN_A^h(z)}{dz} \Bigg{/}
    \frac{1}{N_D^{DIS}}\frac{dN_D^h(z)}{dz}. 
    \label{MultiplicityRatio}	   
\end{align} 
The upper ratio is the number of produced hadrons of species $h$
with energy fraction $z$,  
normalized to the total number of deep inelastic scattering events on
a nuclear target with mass number $A$.
The lower ratio is the corresponding expression for a deuterium target. 
The theoretical calculation of the hadron multiplicity on a nucleus $A$ 
is based on the computation of the hadron survival probability $S_{f,h}^A$
and the fragmentation function  $D_f^h$ in the nucleus:
\begin{align}
  \frac{1}{N_A^{DIS}}\frac{dN_A^h(z)}{dz} =\; &  
       \frac{1}{\sigma^{lA}} \hspace{-0.2cm}
       \int\limits_{\mbox{\footnotesize exp. cuts}}
       \hspace{-0.6cm}
       dx\,d\nu \nonumber\\
   &  \times \sum_f e_f^2 q_f^A(x,\xi_A Q^2) 
      \frac{d\sigma^{lq}}{dx d\nu} S_{f,h}^A(z,\nu) D_f^h(z,\xi_A Q^2).
\label{eq:SIDIS_def}
\end{align}
The total lepton-nucleus DIS cross section $\sigma^{lA}$ is
calculated in leading order  by an integration over the parton
distribution functions  $q_f$ and the lepton-quark cross section
$d\sigma^{lq}/dxd\nu$, including the experimental cuts:
\begin{align}
  \sigma^{lA} & = \hspace{-0.2cm}
    \int\limits_{\mbox{\footnotesize exp. cuts}} 
    \hspace{-0.6cm} 
		 dx d\nu\sum_f e_f^2 q_f^A(x,\xi_A Q^2)
                \frac{d\sigma^{lq}}{dx d\nu} \ .
\label{eq:DIS_def}      
\end{align}
The theoretical calculation in Eqs.~(\ref{eq:SIDIS_def}-\ref{eq:DIS_def}) takes into account several nuclear effects.
The partial deconfinement model \cite{deconfinement} is used to
express the nuclear parton distribution function 
 and the nuclear fragmentation function: both the free-nucleon parton 
 distribution function $q_f(x,Q^2)$ and the free-nucleon fragmentation function
  $D_f^h(z,Q^2)$ are rescaled with a factor $\xi_A=\xi_A(Q^2)$ due to the 
hypothesis that quarks in a bound nucleon have access to a larger
region in space than in free nucleons, i.e.
\begin{align}
 \lambda_A>\lambda_0
\end{align}
where $\lambda_A$ is the confinement scale of a bound nucleon, which is assumed to be proportional to the overlap of nucleons inside
the given nucleus, and $\lambda_0$ is the confinement scale of a free nucleon. 
Effectively, the DGLAP evolution of the nuclear structure functions and the nuclear
fragmentation functions covers a larger interval in momentum compared with the corresponding
functions in the nucleon at the same scale $Q^2$ which implies an increased gluon shower.
Details are presented in \cite{ACC02}.

The hadron survival probability $S_{f,h}^A$ represents   
the probability that the hadron emerges from the nucleus without
having interacted with the nucleus during its evolution 
from the quark to its final state. The survival probability depends on
the struck quark flavor $f$, the hadron species $h$ and the target
nucleus $A$. For simplicity, any scattering of the prehadron with a
target nucleon is assumed to lead to absorption. 
In principle the observed hadron with energy fraction $z$ might have
undergone several inelastic rescatterings before being absorbed, 
whereby its $z$ is degraded. 
However, these multiple scattering processes are suppressed at large
$z$ since the fragmentation function is rapidly falling at $z
\rightarrow 1$, and are expected to be negligible at $z\gtrsim 0.3-0.4$.
In order to calculate $S_{f,h}^A$ 
we modify the simplistic Bialas-Chmaj \cite{BICH83} absorption
formulae used in \refref{ACC02}, which considers the hadronization process to be a decay process, by fully coupling the evolution of the
quark into a prehadron and a hadron with the absorption processes.
The quark decays into the prehadron and has an average lifetime which equals the average formation length 
$\langle l^* \rangle$ of the prehadron after the interaction of the virtual photon $\gamma^*$ with a quark q. 
The prehadron itself decays into the hadron and has an average lifetime
 $\langle \Delta l \rangle$ given by the difference of
the average hadron and the average prehadron formation lengths:
\begin{align}
  \langle \Delta l \rangle = \langle l^h \rangle - \langle l^* \rangle
\end{align}
Here we omit for ease of notation the dependence of the prehadron and
hadron formation length on the struck quark flavor $f$, the hadron
species $h$, the energy fraction $z$ and the virtual photon energy
$\nu$. We will come back later on this topic.
We compute the formation lengths in the framework of 
the standard Lund string fragmentation model.
The Lund model is a semiclassical model which provides a formation length
distribution of the produced prehadrons and hadrons derived from classical 
relations among the production points of the hadrons without quantum fluctuations.  
Since, on the other hand, string breaking is a quantum process akin to
quantum tunneling, we can expect the Lund model to be a good ansatz
for the full probability distribution only up to the first few
moments. For the sake of simplicty, in this paper we consider only the
first moment, i.e., the average formation length.
    
If the initial $\gamma^*q$ interaction occurs at longitudinal
coordinate $y$, 
the probabilities that the intermediate state at time $y'$ is a quark,
$P_{q}(y,y')$, a prehadron, $P_*(y,y')$, or a hadron, $P_h(y,y')$, 
satisfy the following differential equations:
\begin{align}
  \frac{\partial P_{q}(y,y')}{\partial y'}&=-\frac{P_{q}(y,y')}
                                        {\left< l^* \right>}  
                                           \nonumber \\
  \frac{\partial P_*(y,y')}{\partial y'}&=\frac{P_{q}(y,y')}
                                  {\left< l^* \right>} 
                                   -\frac{P_*(y,y')}{\left< \Delta l
				    \right>}
				   -\frac{P_*(y,y')}{\lambda_*(y')}
                                   \nonumber\\
  \frac{\partial P_h(y,y')}{\partial y'}&=\frac{P_*(y,y')}{\left< \Delta l
				    \right>}
				   -\frac{P_h(y,y')}{\lambda_h(y')}
\label{eqn:BCprob}
\end{align}
with initial conditions
\begin{align*}
  P_{q}(y,y'=y) = \,\, & 1 \\
  P_*(y,y'=y) = \,\, & 0  \\
  P_h(y,y'=y) = \,\, & 0 \ .
\end{align*}
The mean free path of the prehadron $\lambda_{*}(y')$
and the  mean free path of the hadron $\lambda_{h}(y')$ are:
\begin{align}
  \lambda_{*,h}(y')&=\frac{1}{A\rho_A(y')\sigma_{*,h}}
\end{align}
where $\rho_A$ is the nuclear density normalized to unity 
and $\sigma_{*,h}$ are the respective cross sections 
of a prehadron and of a hadron. 
The dependence of $\rho_A$  
on the impact parameter $ b$ is suppressed to simplify the notation.
Motivated by the experimental fits in \refref{HERM01,EMC91}
we neglect any final state interaction of the struck quark 
with the nuclear environment (cf. \refref{ACC02}). 
The solutions for the probabilities $P_q,P_*,P_h$ can be  obtained
analytically: 
\begin{align}
 P_q(y,y')&= e^{-\frac{y'-y}{\left< l^* \right>}} \\
 P_*(y,y')&= \int\limits_{y}^{y'}dx\frac{e^{-\frac{x-y}{\left< l^* \right>}}}
            {\left< l^* \right>}
	    e^{-\frac{y'-x}{\left< \Delta l \right>}} 
	    e^{-\sigma_*\int\limits_x^{y'}dsA\rho_A(s)} \\
 P_h(y,y')&= \int\limits_y^{y'}dx'\int\limits_{y}^{x'}dx
            \frac{e^{-\frac{x-y}{\left< l^* \right>}}}
            {\left< l^* \right>} 
	    e^{-\sigma_*\int\limits_x^{x'}dsA\rho_A(s)} 
	    \frac{e^{-\frac{x'-x}{\left< \Delta l \right>}}}
            {\left< \Delta l \right>} 
	    e^{-\sigma_h\int\limits_{x'}^{y'}dsA\rho_A(s)} \ .
 \label{eqn:Ph}
\end{align}
There is a connection between the resulting  
probability distribution $P_q(y,y')$ 
to find a quark at position $y'$ if the initial interaction took
place at $y$ and
the Lund model prehadron formation length distribution. 
The prehadron formation length distribution in our model is simply given
by the first derivative of $P_{q}$ with respect to $y'$ with an opposite sign.
Its first moment equals the first moment of the Lund 
distribution. 
Its second moment deviates from the Lund distribution's second moment
by less than $20\%$ for $z\gtrsim0.5$, with larger deviations at
smaller $z$. An analogous analysis also holds for the hadron formation length
distribution. 
We define the survival probability $S_{f,h}^A$ of the hadron as the
probability that the hadron is not absorbed by the nucleus. It is
obtained as $\lim_{y' \rightarrow \infty} P_h(y,y')$ after integration 
over all $\gamma^*q$ interaction points and impact parameters:
\begin{align}
  S_{f,h}^A = & \int d^2b \int\limits_{-\infty}^{\infty}dy\,\rho_A(b,y)
	      \int\limits_y^{\infty}dx'\int\limits_{y}^{x'}dx 
	      \nonumber  \\
	  & \times \frac{e^{-\frac{x-y}{\left< l^* \right>}}}
              {\left< l^* \right>} e^{-\sigma_*\int\limits_x^{x'}dsA\rho_A(s)}
	      \, \frac{e^{-\frac{x'-x}{\left< \Delta l \right>}}}
              {\left< \Delta l \right>} 
	      e^{-\sigma_h\int\limits_{x'}^{\infty}dsA\rho_A(s)} \ .
  \label{eqn:AbsFac}
\end{align}

\begin{figure}[tb]
\begin{center}
\includegraphics[width=13cm]{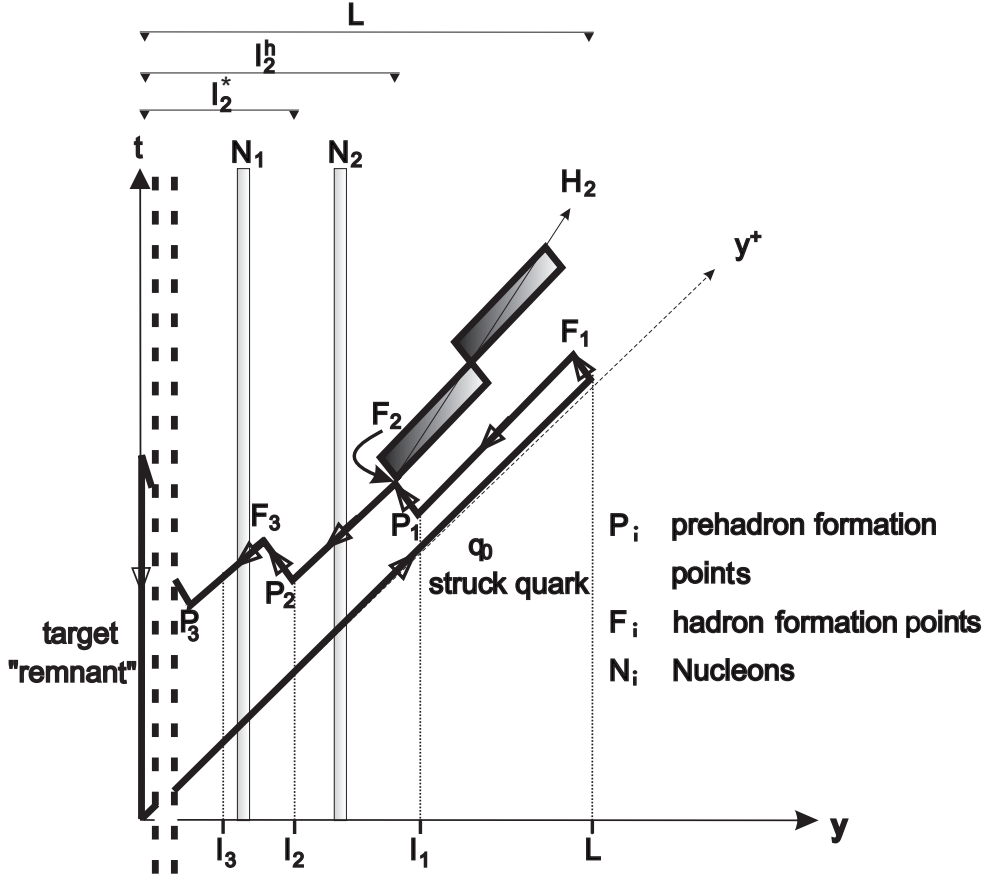}
\caption{Schematic space time picture of hadronization in the Lund model. }
\label{fig:lundmodel}
\end{center}
\end{figure} 

The formation lengths are computed in the framework of the Lund
model~\cite{AND83,BG87}, for which we show a schematic space-time picture 
in  Fig.~\ref{fig:lundmodel}. Due to the relatively high
  measured values of $\nu$, hadronization is supposed to 
occur near the light cone which justifies the neglection of all masses.
This is definitely true for hadrons with large $z$.
The target nucleons $N_1,N_2$ are  at rest in the laboratory frame.
Since the struck quark and the leading hadrons
move close to the light cone, we have enlarged in our illustration 
the negative light cone momenta  and diminished the positive ones for
clarity. 
For the same reason we have omitted a part of the space-time evolution which is 
indicated by the two dashed vertical lines on the left of the figure.
The hadrons $H_i$ are formed at the points  $F_i$ and the index
$i$ gives their rank, counted from right to left.
Each point $P_i$ denotes the production point of one quark anti-quark pair 
due to string breaking. 
We exemplarily consider the production point $P_2$ in the following.
Since the produced antiquark is still connected via a string to the struck quark $q_0$,
we associate with the combined object $(\bar{q}q_0)$ the prehadron $H^*_2$ evolving 
into the hadron $H_2$. Therefore, 
the ''production'' points $P_i$ are the formation points 
of the $i$-th rank prehadron. 
The distance $L$ from the struck quark turning point to the 
$\gamma^*q$ interaction point  sets the scale of hadronization and is
proportional to the energy transferred to the quark:
\begin{equation}
  L=\frac{\nu}{\kappa}
 \label{stringcostant}
\end{equation}
For the calculation in this paper we use the expected vacuum 
string constant \cite{BI80,GY90,CZ92,MAR03} 
\begin{equation}
  \kappa=1 {\rm\ GeV/fm} \ .
\end{equation} 
In order to be consistent with the partial deconfinement model, we rescale the vacuum string
constant in the nucleus. Since the string constant is the physical quantity that sets the confinement
scale, a larger confinement scale corresponds to a smaller string tension in the nucleus:
\begin{align}
 \kappa_A=\frac{\lambda_0^2}{\lambda_A^2}\kappa \ .
\end{align}

\begin{figure}[tb]
\centering
\begin{minipage}{1.0\linewidth}
 \centering
  \begin{minipage}{0.49\linewidth}
   \centering
   \fbox{
   \includegraphics[width=0.85\linewidth]{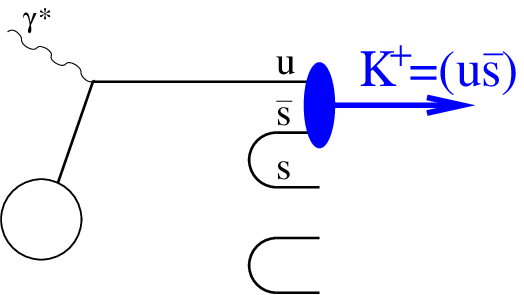}
   }
  \end{minipage}
  \begin{minipage}{0.49\linewidth}
   \centering
   \fbox{
   \includegraphics[width=0.85\linewidth]{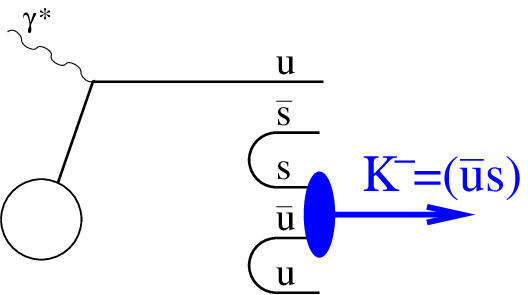}
   }
  \end{minipage}
\end{minipage}
 \label{fig:upK}
\caption{Schematic picture describing the fragmentation
of an up quark into $K^+$ and $K^-$. A similar picture works also 
for fragmentation into $p$ and $\bar{p}$, respectively. }
\end{figure}

In addition, we explicitly consider the dependence of  the formation
time on  hadron species. 
The rank 1 hadron has the struck quark as one of its 
constituents. Therefore, the type of particles which can be produced
as a rank 1 depends  on the flavor of the struck quark.
As an example, in~\figref{fig:upK} we show the different processes
which lead to the production of a $K^+$ and $K^-$ meson from a struck
$u$ quark. The positive kaon can be directly formed at rank 1, whereas
$K^-$ needs a least a  second quark pair produced, i.e. it can only be
produced from rank 2 on. 
More in general,
a negative kaon ($K^-=s \bar u$), as well as an antiproton 
($\bar p = \bar u \bar u \bar d$), cannot be formed as a first rank particle 
by a struck valence quark in the nucleon. As a first rank particle,
the kaon and the antiproton can only be formed from a struck sea
quark, which is a subdominant process at HERMES ($q_{u,d}/q_s \approx
0.05$) included in our calculation. In any case, $K^-$ and $\bar{p}$ can be  
formed from quarks inside the color string ,i.e. as higher rank particles. 
Therefore, kaons and antiprotons are dominantly produced as higher rank particles with shorter formation lengths.
Hence, the Lund string model naturally produces a flavor-dependent
formation time. 
In order to get analytic formulae for the average formation lengths eqns. (\ref{eqn:avgfl1}) 
and (\ref{eqn:avgfl2}) we make several approximations:
we compute the average formation time 
with the standard Lund string fragmentation function $f(u)\propto(1-u)^{D_a}$ 
(cf. \refref{ACC02}) which does not depend on the mass of the produced hadrons.
We do not take into account the hadron masses in the determination of the relation
between the hadron energy and the hadron production point. 
The main effect of an inclusion of hadron masses 
in the calculation is that the average formation length is cut off 
at small values of z due to energy conservation
but that the general behavior at larger $z$ is not significantly affected.  
We average over an infinite number of produced
hadron ranks. 
Furhermore, we do not take into account gluonic 
excitations of the strings and neglect transverse momenta in the
 computation of the average formation time.
For details about the derivation of the average formation times we refer 
to \refref{ACC02}. 
Here we only show how the mean formation time of a hadron
which can be produced  from all ranks $n \ge 1$ differs from the
formation time of a hadron which  
can only be produced from rank 2 on, i.e. $n \ge 2$. The average
prehadron formation length for hadrons which are producible as first
rank particle reads: 
\be
  \langle{l_{\ge 1}^*}\rangle  &=& \frac{1+D_a}{1+C+(D_a-C)z}\ (1-z)\, z\, L
  \nonumber \\  
        & &\left[ 1 + \frac{1+C}{2+D_a}\,\frac{(1-z)}{z^{2+D_a}}\,
       {}_2F_1\Big( 2+D_a,2+D_a;3+D_a;\frac{z-1}{z} \Big)\right] 
       \label{eqn:avgfl1} 
\ee
and the average prehadron formation length for particles which are not producible as
a first rank particle is given by:
\be
  \langle{l_{\ge 2}^*}\rangle  &=& (1-z)\, z\, L
  \nonumber \\  
        & &\left[\frac{1+D_a}{2+D_a}\,\frac{1}{z^{2+D_a}}\,
       {}_2F_1\Big( 2+D_a,2+D_a;3+D_a;\frac{z-1}{z} \Big)\right]
       \label{eqn:avgfl2}  
\ee  
Here ${}_2F_1$ is the Gauss hypergeometric function, the parameters
$C$ and $D_a$ with $a=(q,qq)$ arise from the string 
fragmentation function. Furthermore, these parameters select whether a meson
($a=q$) or a baryon ($a=qq$) is produced (see \refref{ACC02}). Their
numerical values are given in Ref. \cite{AND83} by $C=D_q=0.3$ and
$D_{qq}=1.3$. 
There is a simple rank-independent connection between the prehadron
and the hadron formation length
for fixed  energy fraction $z$ carried by the hadron :
\begin{align}
 \langle{l_{\ge 1,2}^h} \rangle = \langle{l_{\ge 1,2}^*}\rangle +
   z L \ .
\end{align}

\begin{figure}[tb]
\vskip-4.5cm
\centering
\begin{minipage}{1.4\linewidth}
 \hspace{-1.5cm}
 \begin{minipage}{0.49\linewidth}
  \includegraphics[width=1.0\linewidth]{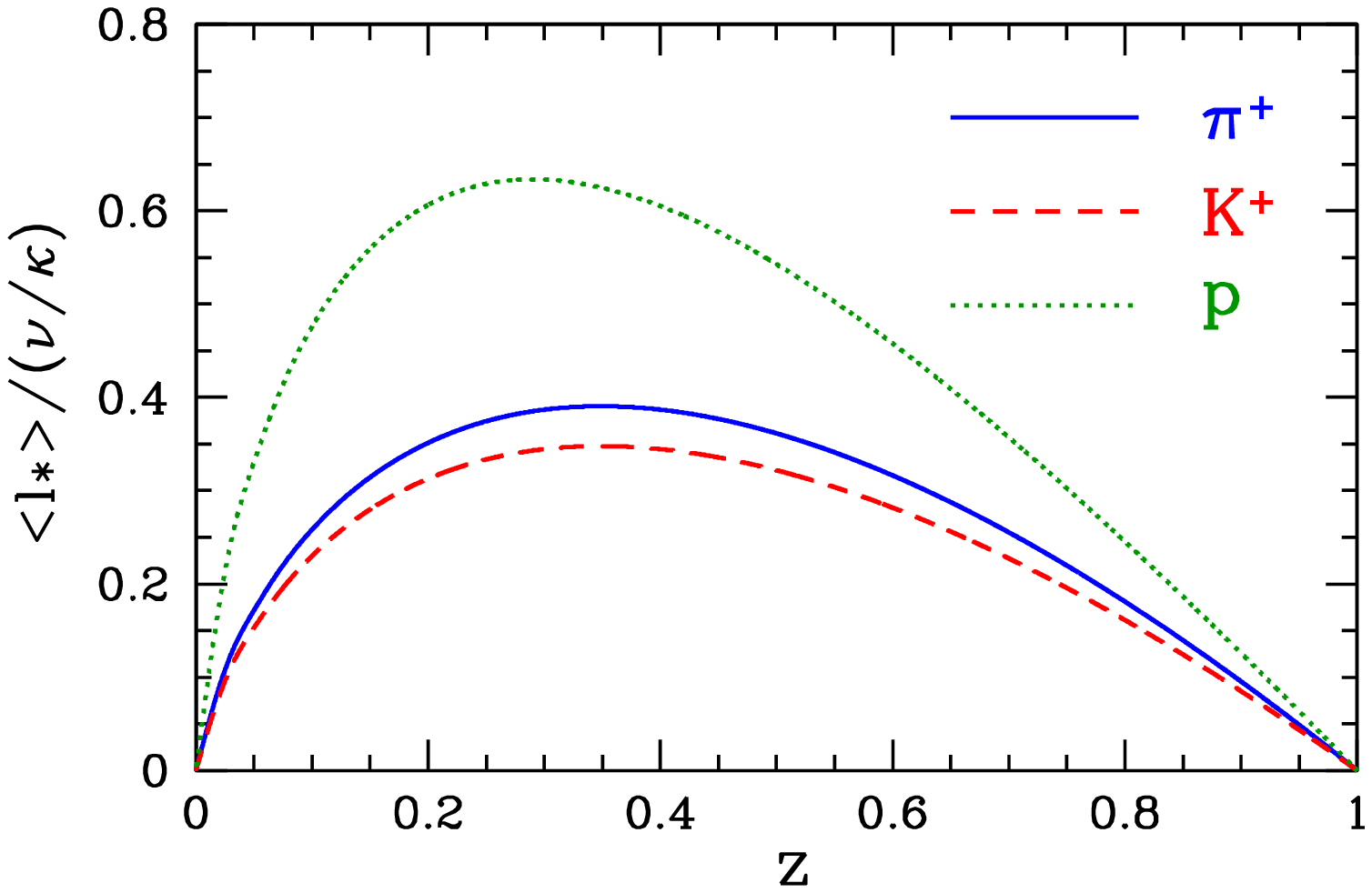}
 \end{minipage}
 \hspace{-2.5cm}
 \begin{minipage}{0.49\linewidth}
  \includegraphics[width=1.0\linewidth]{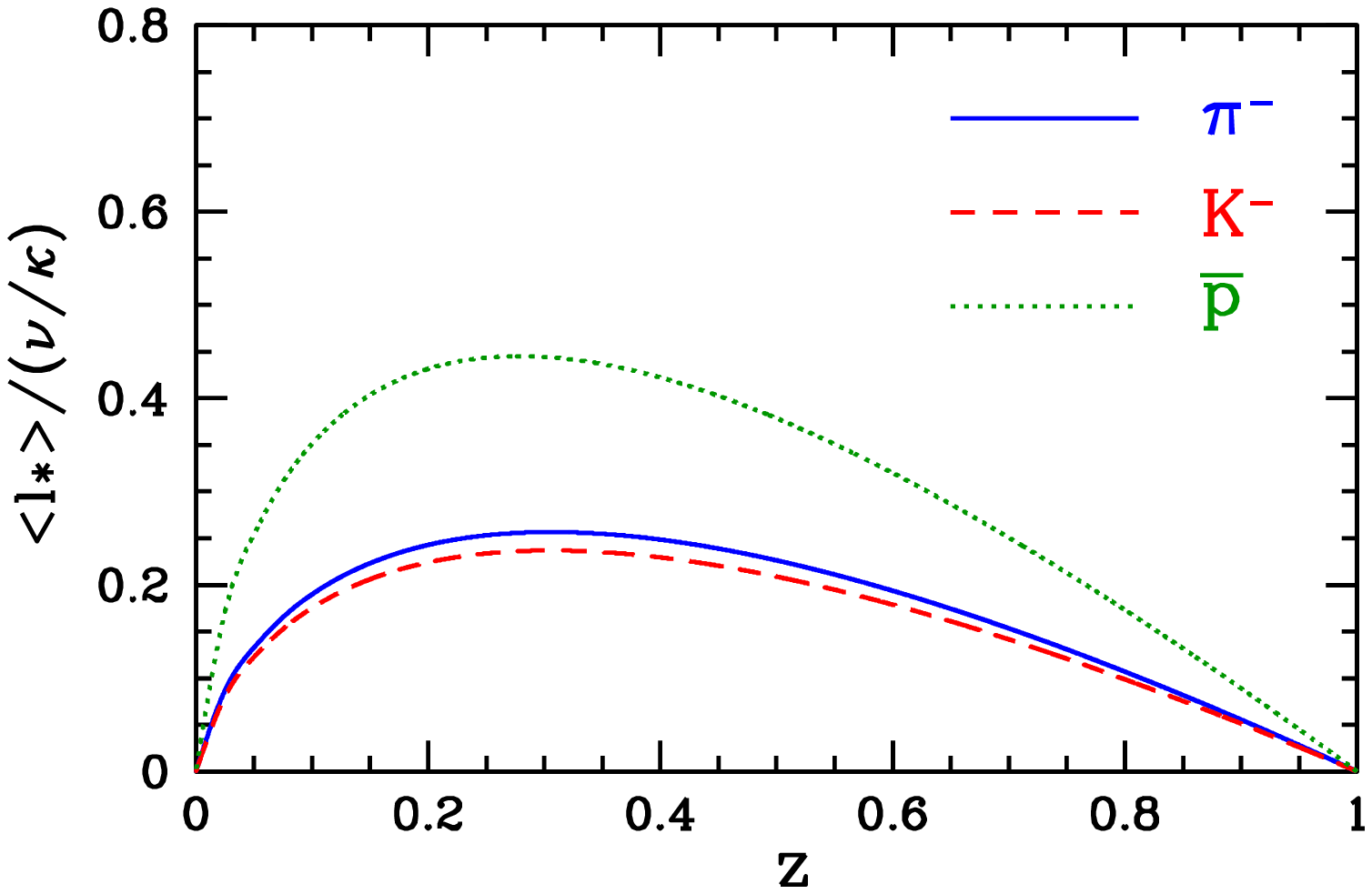}
 \end{minipage}
\end{minipage} 
\vskip-0.5cm 
\caption{Computed prehadron formation lengths when an up quark is
  struck by the virtual photon. {\it Left:} When a $\pi^+$, $K^+$ or
  $p$ is observed, the corresponding prehadron can be created at 
  rank $n \ge 1$. {\it Right:} When a $\pi^-$, $K^-$ or
  $\bar p$ is observed, the corresponding prehadron can be created
  only at rank $n \ge 2$. }
\label{fig:rank12}
\end{figure}

The present model is based on string fragmentation in (1+1)-dimensions.
Realistic hadrons also have a transverse extension. 
We expect that 
the prehadron does not yet have the full size of the hadron and
therefore interacts with a smaller cross section, 
$\sigma_*< \sigma_h$. 
We fit the prehadronic cross section to reproduce the
pion multiplicity ratio on ${}^{84}$Kr \cite{HERM03} and we obtain:
\begin{equation}
  \sigma_*=\frac{2}{3} \sigma_h \ .
\end{equation}
Such a prehadronic cross-section is in agreement with 
\refref{FALTER04}, where the multiplicity ratio of charged hadrons 
for the HERMES experiment is computed with a prehadronic
cross-section 
increasing quadratically during the hadron formation time
$\tau_f=0.5 {\rm\ fm}/c$ 
from $0.3\sigma_h$ to the full hadronic cross-section. 
On the average one obtains a prehadronic cross-section 
$\langle\sigma_*\rangle=0.63\sigma_h$.
We use the proportionality factor $2/3$ for all other hadronic species.
Empirically we know that baryons have a larger radius than $\pi$ mesons,
which are larger than $K$ mesons. 
We renormalize the string tension, i.e. the quantity which
sets the confinement scale to the
confinement scales $r_h^2$ of real hadrons taken from
Ref. \cite{Radii}  
\begin{equation}
\kappa_{h}=\kappa \frac{r_{\pi}^2}{r_{h}^2}.
\end{equation}
This produces an increased formation length for baryons due to the larger
size of the proton compared with pions.
A comparison between the computed formation lengths of pions, kaons,
protons and antiprotons is shown in Fig.~\ref{fig:rank12}.  
The flavor dependence due to the mechanism exemplified in Fig.~\ref{fig:upK} decreases 
the average formation length of the negative mesons by about $30\%$
compared to the positive ones, and by more than $40\%$ for $\bar{p}$
compared to $p$. 
The effect of the baryon's larger size results in a slightly longer baryon
formation time compared with the meson ones.

\begin{figure}[tb]
\vskip-3.5cm
\centering
\begin{minipage}{1.3\linewidth}
 \hspace{-1.2cm}
 \begin{minipage}{0.49\linewidth}
  \includegraphics[width=1.0\linewidth]{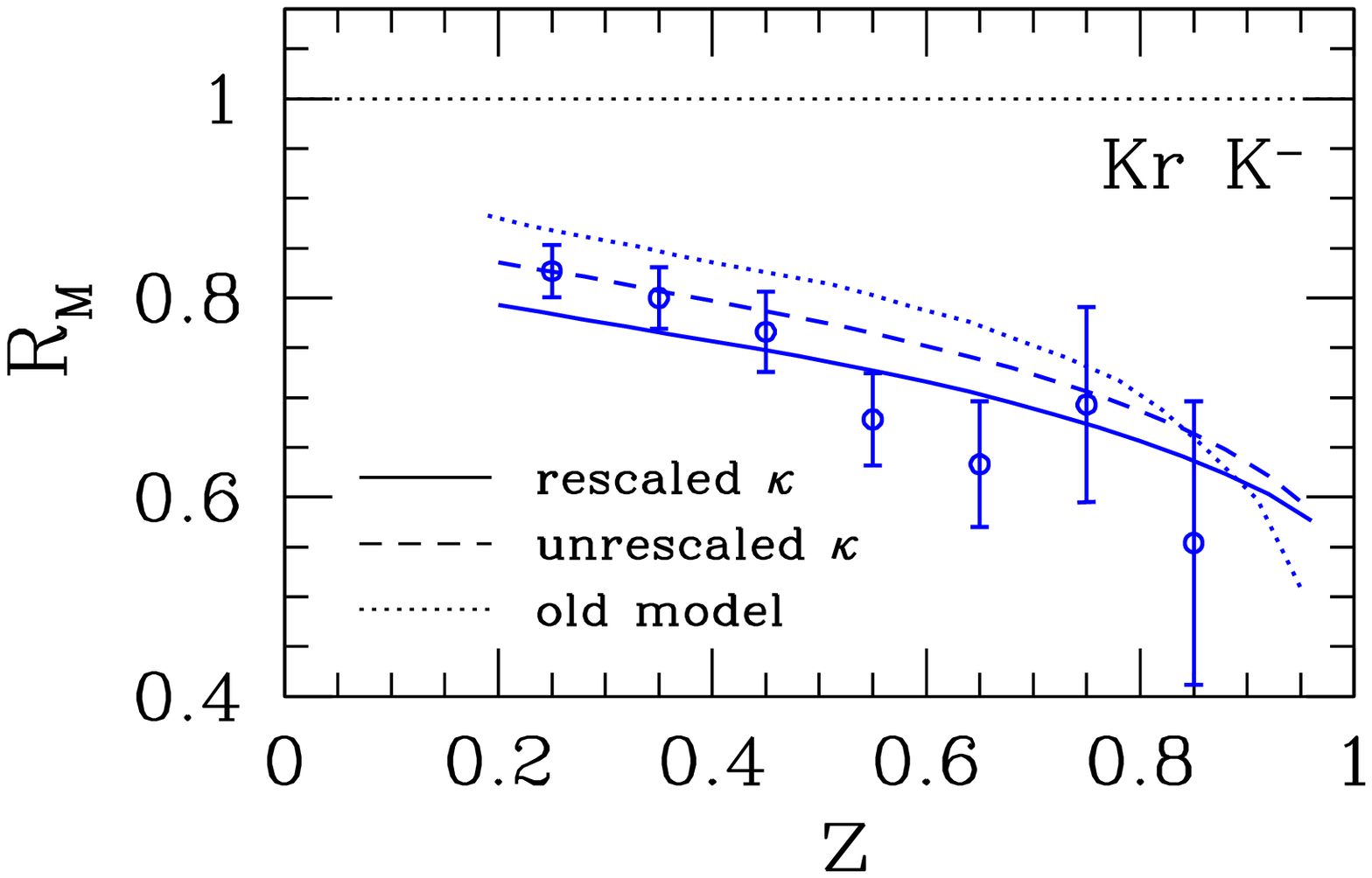}
 \end{minipage}
 \hspace{-1.5cm}
 \begin{minipage}{0.49\linewidth}
  \includegraphics[width=1.0\linewidth]{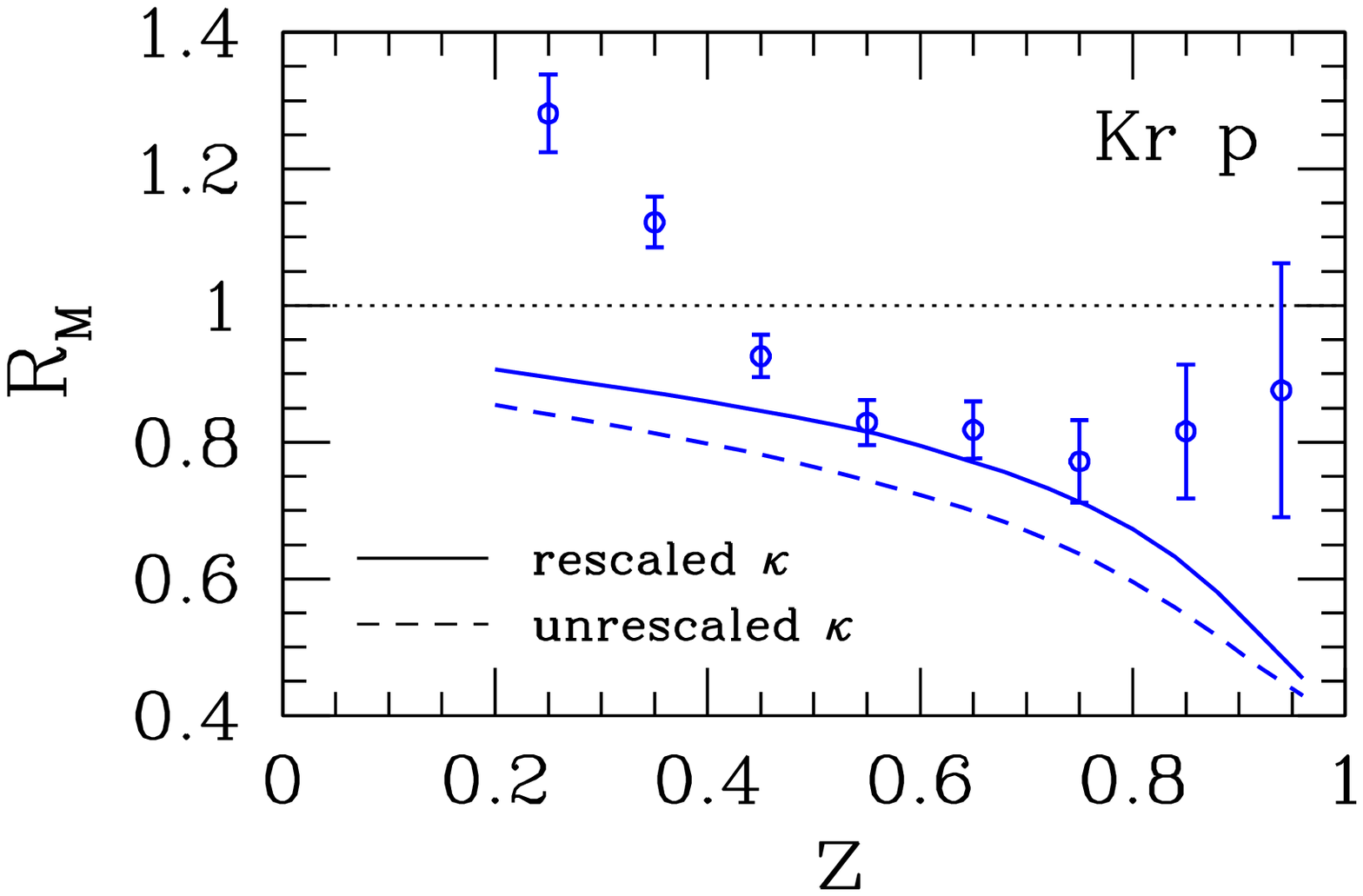}
 \end{minipage}
\end{minipage}  
\caption{Multiplicity ratios for  $K^-$ and $p$ on krypton are shown  as function of
 $z$ in the left  and right panel respectively. Solid (dashed) curves
are computed with (without)  hadron size rescaling in the formation
 length. The dotted curve shows the published result~\cite{ACC02} for $K^-$ without the above described model modifications. The data are taken from Ref.~\cite{HERM03}.}
\label{fig:RM-for-Km-prot}
\vskip.4cm
\end{figure}

The effect of the modified string tension $\kappa_h$ on the  $K^-$ and $p$
multiplicity ratios for krypton is shown  in the left and right parts of
Fig.~\ref{fig:RM-for-Km-prot}, respectively. 
The dashed line represents the computation with the unrescaled string tension
$\kappa$, while the solid line shows the effect on
the multiplicity ratio of the increased (reduced) string tension $\kappa_h$ for kaons (protons).
Thus the modification of the string tension by accounting for hadronic size  
improves
the agreement between the computed and measured multiplicity ratios 
both for  $K^-$  and $p$.
In addition, the published 
results~\cite{ACC02}  for  $K^-$  without any of the  above described model modifications 
are shown  as dotted line.
Including in the model   a flavor dependent formation length, 
the realistic
 string constant and  the reduced prehadron cross section improves 
our previous computation~\cite{ACC02}. The proton spectra, however,
cannot be described by the presented model adequately. We'll discuss
 this issue in \secref{sec:multrat}.
 
Diffractively produced vector mesons have typically large $z$ and may influence the 
multiplicity spectra at intermediate $z$ by rescattering prior to their 
decay, cf. \refref{FALTER04}. A recent experimental analysis in
\refref{HERM03} shows that diffractively produced vector mesons affect
the pion multiplicities by a few percent only and therefore are neglected
in our computation for all meson species.


\section{Mass-number dependence}
\label{sec:A-dependence}

The dependence of hadron attenuation on the mass number $A$ of 
the target nucleus is commonly believed to clearly distinguish the
absorption and energy loss mechanisms for hadron attenuation. Indeed,
the common expectation is that attenuation in absorption models is
proportional to the hadron in-medium path length $L$, leading to 
$1-R_M \propto A^{1/3}$. On the other hand, the average energy loss for
a parton traversing a QCD medium is proportional to $L^2$, which leads to $1-R_M
\propto A^{2/3}$. 
In the following we will discuss in detail the $A$-dependence for the  absorption  model with 
the main goal to obtain insights via analytic formulae. A numerical analysis of the $A$-dependence
is presented in Section~\ref{sec:results}.

For illustrative purposes and to obtain analytic formulae we take the
model described \secref{sec:DevModel} in 
the one step
approximation where we neglect final hadron production in the
calculation of the survival probability, 
since  at $z\ge0.2$ the average hadron
formation length 
is longer than 10 fm, as shown in
Table~\ref{table:c1c2}.
To further simplify, we consider the case of a hard-sphere nucleus of
mass number $A$ and radius $R =  r_0 A^{1/3}$, with $r_0=1.12$ fm \cite{PovhBook}. We
also consider only hadronization from rank 1 on, dominant in the HERMES
kinematics except for $K^-$ or $\bar p$ production. Neglecting
absorption in deuterium and the small rescaling correction we find
that the hadron multiplicity ratio $R_M$ equals the hadron survival
probability $S_A$. 
Therefore, the hadron attenuation $1-R_M$ can be
  approximated as:  
\begin{align}
  1-R_M \approx \, & 1-S_A  \nonumber \\
  = \, & 1-\frac{\pi
    \rho_0}{A}\int\limits_0^{R^2} db^2\int\limits_{-R(b)}^{R(b)}dy
    \int\limits_{y}^{\infty}dx\,\frac{e^{-\frac{x-y}{\left< l^* \right>}}}
            {\left< l^* \right>}
	    e^{-\rho_0\sigma_*\int\limits_x^{\infty}ds
	     \Theta(R(b)-|s|)} \ ,
 \label{1-rm}
\end{align}
where we omitted all flavor and hadron species dependence
  for simplicity. In \eqeqref{1-rm} 
$\rho_0$ is the nuclear density, $r_0 =
\left(\frac{3}{4\pi\rho_0}\right)^{1/3}$ and  
$R(b)=\sqrt{R^2-b^2}$. 
With the mean free path of the prehadron
$\lambda_*=1/(\sigma_*\rho_0)$ 
the expression for the attenuation simplifies:
\begin{align}
  1-R_M=\frac{\pi \rho_0}{2 A} \langle l^* \rangle^3
    \int\limits_0^{2R/\langle l^* \rangle} dt\,t
    \int\limits_0^t dr \int\limits_0^r du\, e^{-u}
    \left[ 1-e^{\frac{\langle l^* \rangle}{\lambda_*} (u-r)} \right]
 \label{BCsimple}
\end{align}
The attenuation is controlled by two ratios; firstly,
$a=\vev{l^*}/\lambda_*$, the ratio of formation length to mean free
path and, secondly, $b=2R/\vev{l^*}$, the size of the 
nucleus relative to the formation length.  If 
both quantities are small one can expand the integrand 
in powers of $u$, obtaining after all the integrations
a series which starts with $b^2 \propto A^{2/3}$,
contrary to common expectations which dictate $A^{1/3}$:
\begin{align}
  1-R_M=\frac{1}{10}a b^2 - \frac{1}{48} (1+a)a b^3 
    + \frac{1}{280}(1+a+a^2) a b^4 + \mathcal{O}[b^5] \ .
\end{align}
However, since $a$ and $b$ are of order unity for medium to
heavy nuclei, the series converges very slowly, we cannot state
whether or not it is possible to approximate the full expression 
by the leading term $\propto A^{2/3}$. To improve the convergence
we fit the $r$-dependence of the integrand to a numerical function
with a simpler form:
\begin{align}
  \int\limits_0^r du e^{-u}\left[ 1-e^{a(u-r)} \right]
= \frac{1-e^{-ar}-a\left( 1-e^{-r} \right)}{1-a}  
\approx 1-e^{-w a r^2} \ .
  \label{approxint}
\end{align}
A good fit can be achieved with a constant
\begin{align}
  w=0.19
\end{align}
independently of $r$ and $a=\frac{\vev{l^*}}{\lambda_*}$ in the range
$0<r<5$, $0<a<3.5$  where the function differs substantially from zero. 
Using \eqeqref{approxint} in \eqref{BCsimple}, one finds a rapidly 
converging expansion of the attenuation, since now the small 
$w$-parameter enhances the convergence:  
\begin{align}
  1-R_M & =  \frac{1}{5} w a b^2  - \frac{3}{70} (w a b^2)^2 
    + \mathcal{O}[(wab^2)^3] \nonumber \\
  & = c_1 A^{2/3}+c_2 A^{4/3} + \mathcal{O}[A^2]  \ .
 \label{form:survBC}
\end{align}

\begin{table}[tb]
\begin{center}
\begin{tabular}{|c|c|c|c|c|} \hline
$z$ & $\langle l^h(z) \rangle$ [fm]& $c_1$  & $c_2$     & $\bar A$
\\ \hline \hline
.25 & 10.15                        & 0.0095 & -0.000096 & 980 \\ \hline
.45 & 11.72                        & 0.0103 & -0.000114 & 860 \\ \hline
.65 & 12.34                        & 0.0142 & -0.000217 & 530 \\ \hline
.85 & 11.98                        & 0.0314 & -0.001059 & 160 \\ \hline
\end{tabular}
\end{center}
\caption{Average hadron formation time and values of the $c_1$ and
  $c_2$ coefficients in \eqeqref{form:survBC} for $\pi^+$ production
  at different $z$ values. The mass number  $\bar A$ at which 
  $A^{4/3}$ terms in \eqref{form:survBC} become comparable with
  $A^{2/3}$ terms are also given. For each value of $z$, we
  have taken the appropriate average value of $\nu$ measured at 
  \cite{HERM03}. The large value of $\vev{l^h}>10$ fm justifies
  neglecting hadron interactions with the nucleus.}  
\label{table:c1c2}
\vskip.3cm
\end{table} 

We have computed the coefficients $c_1$ and $c_2$ in Eq.~(\ref{form:survBC})
for $\pi^+$-production and different values of $z$. The results are
presented in Table~\ref{table:c1c2}. 
We observe that the series  converges quickly over the whole $z$
range. Therefore, it makes sense to approximate the nuclear attenuation with a
power law,
\begin{align}
  1-R_M = c A^\alpha \ ,
 \label{powerlawfit}
\end{align}
where 
\begin{align}
  c & \ \approx \ c_1(z) \\ 
  \alpha & \ \approx \ 2/3 \ 
\end{align}
since higher-order terms will not give large corrections except at large $z \gtrsim 0.8$. Indeed from Eq.\eqref{form:survBC}, we
see that higher order terms are negligible if
\begin{align}
  A \ll \left| \frac{c_1}{c_2} \right|^{3/2} \equiv \bar{A}  \ .
\end{align}
Values of $\bar A$ are given in \tabref{table:c1c2}, and become
comparable to mass numbers of medium-heavy nuclei  around $z=0.8$. We
will further discuss the power-law fit \eqref{powerlawfit} in
Section~\ref{sec:results}.   

We conclude with three remarks. First, a lot of information on
absorption dynamics is contained in the proportionality coefficient $c$
of Eq.\eqref{powerlawfit}, and its strong 
dependence  on $z$ needs to be taken into account when analyzing
experimental data and theoretical models. 
Second, the approximate power law $1-R_M \propto
A^{2/3}$ is not a peculiarity of the absorption model we developed,
but has a more general validity. Indeed, it is possible to show that it
appears whenever the probability distribution of the prehadron
formation length $\Delta y$ is analytic and has a finite limit as
$\Delta y \ra 0$. The origin of the $A^{2/3}$ is geometric and related
to the integration over the $\gamma^*$-quark interaction
points. Non-analytic probability distributions may give leading powers
different from 2/3 (cf. appendix \ref{app:genericabs}).
Clearly, the calculation based on a hard sphere nucleus is not
adequate for light nuclei. In the following numerical
calculations of the full model discussed in Sect.~\ref{sec:DevModel} we
use appropriate density profiles for all nuclei.


\section{Numerical results}
\label{sec:results}

\subsection{Multiplicity ratio for $\pi$, $K$, $p$ and $\bar{p}$ }
\label{sec:multrat}

The recent HERMES data have shown significant differences in the
multiplicity ratios of various hadrons. 
As suggested in Refs.~\cite{KOP03,WW02,arleo02} 
the observed flavor-dependence of the multiplicity ratio may be useful to disentangle  
different theoretical frameworks describing the hadronization process.
However, the aforementioned   models provide
computations for nuclear attenuation of charged hadrons~\cite{WW02},
charged pions and kaons \cite{KOP03,arleo02,AK02} only. 
Theoretical computations for protons and antiprotons, that are expected to
be very sensitive to the formation length mechanism as shown
in  Fig.~\ref{fig:rank12}, are given in \refref{FAL03}. 
These  predictions
generally  underestimate  baryon
multiplicity ratio, especially for the krypton nucleus and for $z<0.4$.
 
\begin{figure}[tb]
\begin{center}
\hspace*{-4.4cm}
\includegraphics[width=18.0cm, height=17.5cm]{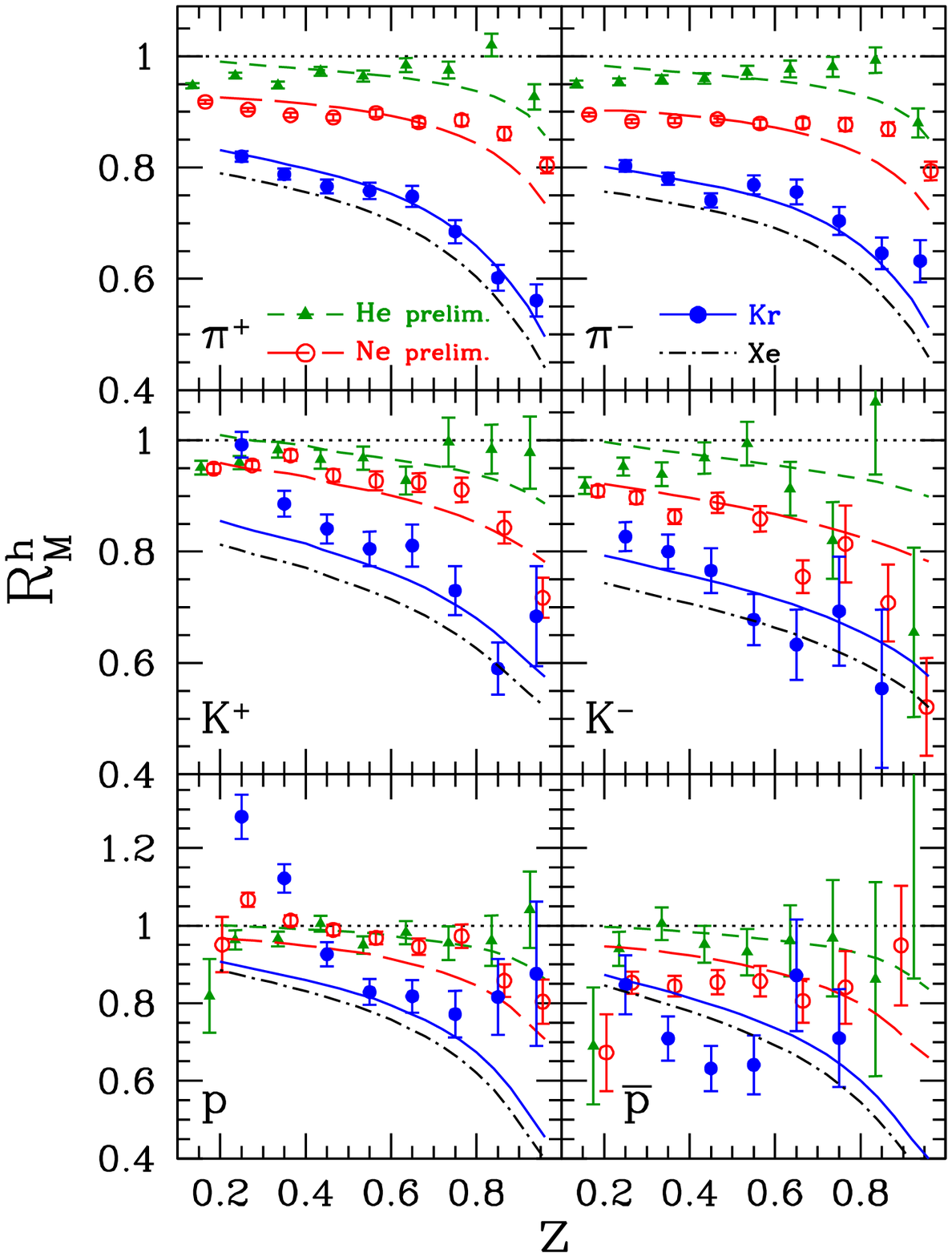}
\caption{Computed multiplicity ratios (cf. Eq.~\ref{MultiplicityRatio}) for  pions, kaons, protons and
   antiprotons  as a function of $z$ for He (dashed line),
   Ne (long-dashed line), Kr (solid line) and Xe (dot-dashed
   line). The HERMES data on Kr (solid circles)~\cite{HERM03} and
the preliminary data on He (closed triangles) and on  Ne (open circles)
~\cite{HERM04} are shown with their statistical uncertainties.
 To improve the readability of the figure,  He and Ne
   experimental points have been shifted by 0.015 to the left and to the
   right, respectively.} 
\label{fig:RM-vs-z}
\end{center}
\vskip.5cm
\end{figure}

\begin{figure}[tb]
\begin{center}
\hspace*{-4.4cm}
\includegraphics[width=18.0cm, height=17.5cm]{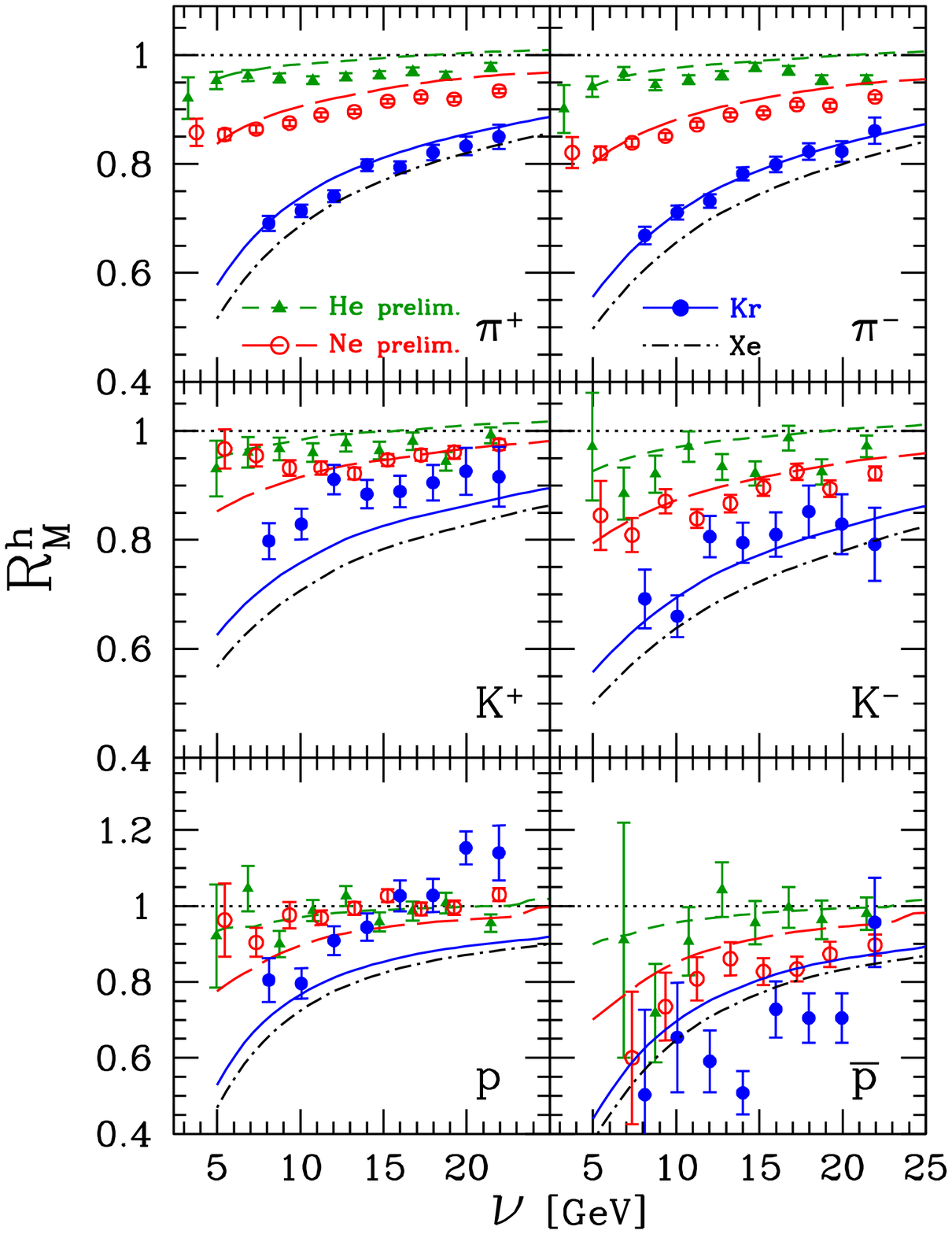}
\caption{Computed multiplicity ratios (cf. Eq.~\ref{MultiplicityRatio}) for  pions, kaons, protons and
   antiprotons  as a function of $\nu$ for He (dashed line),
   Ne (long-dashed line), Kr (solid line) and Xe (dot-dashed
   line). The HERMES data on Kr (solid circles)~\cite{HERM03} and
the preliminary data on He (closed triangles) and on  Ne (open circles)
~\cite{HERM04} are shown with their statistical uncertainties.
To improve the readability of the figure, He and Ne
   experimental points have been shifted by 0.25 GeV to the left and to the
   right, respectively.} 
\label{fig:RM-vs-nu}
\end{center}
\end{figure}

In order to have charge-separated fragmentation functions,  we use
the leading order Kretzer parametrization \cite{KRE00}
of the fragmentation functions for pions and kaons, while for protons
and antiprotons we use the Kniehl-Kramer-Potter \cite{KKP00}
parametrization.
Predictions of our model for $\pi^\pm$, $K^\pm$, $p$ and $\bar{p}$
multiplicity ratios on He,  Ne, Kr and   Xe  
are shown for the $z$ (Fig.~\ref{fig:RM-vs-z}) and $\nu$
(Fig.~\ref{fig:RM-vs-nu}) dependence, and compared to published HERMES
data for identified hadrons on Kr~\cite{HERM03} and to
preliminary data on He~\cite{HERM04} and  Ne~\cite{HERM04}. (We do not include computations and final data on N as they are very close to Ne.)

The $z$ dependence reported in Fig.~\ref{fig:RM-vs-z} shows nice
agreement with the data 
for  negative and positive pions for all the nuclei.
The predictions for $K^+$ and $K^-$ are in qualitative agreement with
the trend shown by the data,
except for the low-$z$ region. This region contains contributions from
both leading hadrons decelerated 
in nuclear rescattering and from secondary $K^+$ and $K^-$ produced from 
initial pions and $\rho$-mesons that are more abundant than strange particles.
Both these  contributions are not accounted in  the purely absorption
treatment of the final state interaction included in our model. 
The experimental $z$ dependence of the antiprotons data is qualitatively
reproduced within statistical uncertainties by the model, while the
proton multiplicity ratio is underestimated 
especially in the low $z$ region.
The discrepancy at low-$z$  repeats itself in the $\nu$-dependence shown in
Fig.~\ref{fig:RM-vs-nu}, because an average $z$=0.3-0.4 enters in the
$\nu$-dependence.
A simple explanation in terms of a smaller prebaryonic cross-section or a larger baryon formation is clearly insufficient to account for the large rise of the proton multiplicity ratio $R^p_M$ above 1 at $z\lesssim0.4$: indeed nuclear absorption can only reduce $R_M$ below 1 and rescaling effects, which can increase $R_M$ above 1 at low enough $z$, are too weak to describe the observed data. Another possible explanation for the difference in the proton sector
is that theoretical computations do not account for the $z$
degradation due to (pre)hadron rescattering nor final state
interactions and decays. However, even a full transport model like
\refref{FAL03,FALTER04} which takes these effects fully into
account, fails to reproduce the rapid growth of the proton data at low
$z$. In addition, the contribution of protons from target
fragmentation is ruled out as an explanation of the effect, because
the experimental momentum cut ($|\vec{p}_N|>4$ GeV) 
removes most of the protons knocked out of the nucleus.
The discrepancy in the low-$z$ region may also point to a non negligible
diquark contribution as suggested in \refref{FALTER04}.
 

\subsection{Mass-number dependence: a new analysis}

The simplest way of analyzing the $A$-dependence of 
hadron attenuation  $1-R_M$ is in terms
of a power law:
\begin{align}
  1-R_M^h(z;A) = c^h(z;A) \, A^{\alpha^h(z;A)} \ .
 \label{power-law}
\end{align}
For simplicity, we consider only hadron
attenuation in fixed $z$ bins, integrated over $\nu$ and $Q^2$; a
similar discussion applies for arbitrary bins in $z$, $\nu$ and
$Q^2$. The coefficients $c$ and $\alpha$ both depend in general on
$z$ and $A$. Indeed, as
we have seen in section 3, the absorption model gives
a power series:  the lowest order term is
$A^{2/3}$, but higher order terms become important for large $A$ and $z$.
Moreover, 
since we expect a strong dependence of the coefficient $c$ on $z$ 
a single constant $c$ cannot fit all experimental bins.
Therefore, we propose to analyze the  $A$-dependence of our model and
experimental  data alike, by fitting the hadron attenuation to the
power law \eqref{power-law} 
leaving both $\alpha$ and $c$ as free parameters.  
These parameters are strongly correlated: indeed a small increase
of the exponent $\alpha$ can be compensated by a small decrease of the
coefficient $c$, and vice-versa. Therefore the fit results 
are best presented as $\chi^2$ contour plots in the $(c,\alpha)$-plane,
showing the aforementioned correlations.  

The best-fit parameters $c$ and $\alpha$ are determined by chi-square
minimization, i.e. the $\chi^2$ merit function  
\begin{align}
  \chi^2(c,\alpha) = \sum\limits_i
    \frac{1}{\sigma_i^2}[(1-R_M^h)(A_i)-c A_i^\alpha]^2 
\end{align}
is minimized with respect to $c$ and $\alpha$. Here $\sigma_i$ is the
uncertainty of the theoretical or the experimental  points
respectively. 
  For the experimental data we use the statistical uncertainty
  only, while in the case of the theoretical computations we use the
  quadratic sum of the uncertainties corresponding to the precision of the
  numerical computation ($\approx$1\%), and to the choice of model parameters.
  In our model we fit the value of the prehadron-nucleon 
  cross section $\sigma_*$ to experimental data for $\pi^+$
  production on  Kr. Hence the theoretical relative uncertainty is
  determined by the experimental relative uncertainty of $\pi^+$
  attenuation on Kr, which approximately equals $6\%$ independent of $z$.
  Then, we assume that 
  the uncertainty of $\sigma_*$ similarly yields a $6\%$ relative uncertainty on
  hadron attenuation independently of $z$ and $A$, as well. A change in the
  other parameters of the model would be reflected in a slightly different
  value of $\sigma_*$, but would not affect the estimate of the
  theoretical uncertainty of hadron attenuation.


The $\chi^2$ contour plots in the
($c,\alpha$)-space are computed as constant $\chi^2$
boundaries, enclosing a region such that
\begin{align}
  \chi^2_{min}<\chi^2< \chi^2_{min}+\Delta \chi^2
\end{align}
where $\chi^2_{min}$ is the minimum of the $\chi^2$ function, obtained
at the best-fit parameters. We fix $\Delta \chi^2=4.61$ which
corresponds, in the case 
of normally distributed fit parameters, to the region which covers
$90\%$ of the total probability distribution. 

\begin{figure}[ht]
\centering
\hspace*{-2.25cm}
\includegraphics[width=18.0cm, height=19.cm]{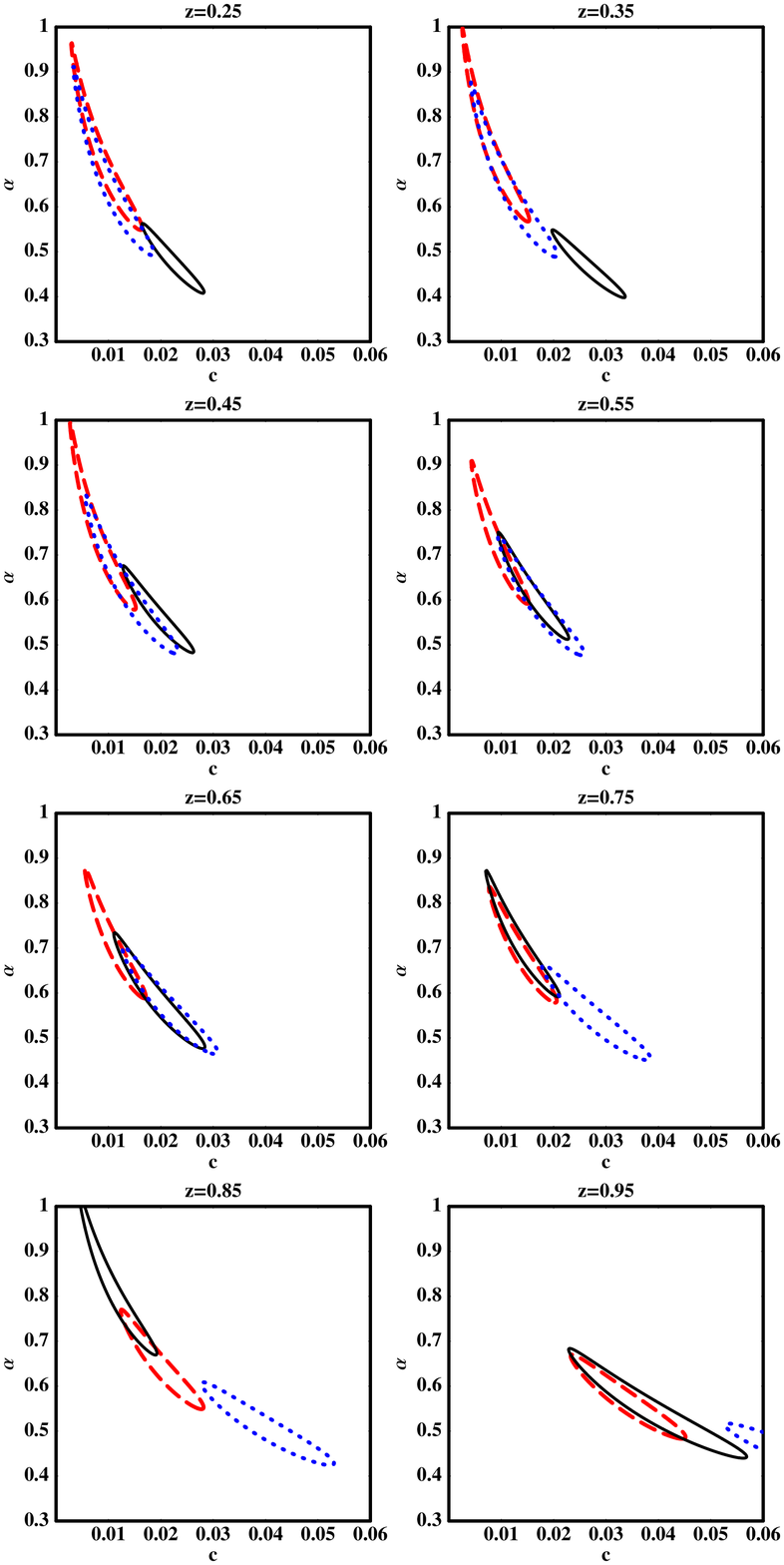}
\vskip-0.5cm
\caption{$\chi^2$ contours for the fit $1-R_M=c(\nu,z,h)\,A^\alpha$
  on  $^{4}$He,  ($^{14}$N), $^{20}$Ne and  $^{84}$Kr nuclei, in fixed
  $z$-bins. The fits to our model computation are shown by the 
  dashed $\chi^2$ contours in the case of the pure absorption model and by dotted $\chi^2$ contours in the case of the full model (absorption + rescaling). The solid contours show the fits to the HERMES  pion multiplicity ratios on  (N~\cite{HERM01}),
  Kr~\cite{HERM03} and to the preliminary data on   He and  Ne
  \cite{HERM04}, accounting for statistical uncertainties only. The
  parenthesis on (N) indicate that nitrogen is included in the fit
  only for $z\ge0.55$. Note that in the last $z$ bin the contour for the full theory computation is partially out of the plot range.} 
\label{fig:fig6}
\end{figure}

The dotted $\chi^2$ contours in Fig.~\ref{fig:fig6} represent 
the fit to the numerical computation in various $z$-bins of the full
theoretical model as it is described in section 2, including $^{4}$He,
($^{14}$N), $^{20}$Ne and $^{84}$Kr nuclei. In order to have a direct
comparison with the theoretical mass number dependence 
analysis of the pure absorption model in section 3, the rescaling of
parton distribution and fragmentation functions is not included 
in the numerical computation yielding the dashed contours. 
The solid $\chi^2$ contours are obtained from the published HERMES data on
($^{14}$N~\cite{HERM01}) and $^{84}$Kr~\cite{HERM03} and the
preliminary data on $^{4}$He~\cite{HERM04} and
$^{20}$Ne~\cite{HERM04}. The notation ($^{14}$N) indicates that this
nucleus is included in the fits of theory and experiment only for
$z\geq 0.55$, because there are no experimental data in lower $z$-bins. The fits at $z\le0.45$ contain only three data points, namely He, Ne and Kr, 
and 2 fit parameters. Therefore, the fit yields in these bins at best a rough
estimate only. Hence, we consider the results in these $z$ bins to be
less reliable. 

The presented results  on  He, (N), Ne and Kr  at $z\ge0.55$ for 
the pure absorption model computation are
qualitatively in agreement with the trend shown by the experimental
data, except for the $z=0.85$ bin, where the experimental fit gives
the largest $\alpha$ due to $1-R_M<0$ value for $^{4}$He.
At small $z$, the full model contours behave similarly to the pure absorption contours, but the two become more and more separated as $z$ is increased.
Furthermore, the full model increasingly disagrees with experimental data when $z\geq0.75$.
This shows the power of the proposed kind of analysis, which is
a possible tool to disentangle different theoretical models.  

The slope of the $\chi^2$ contour plots shows a trend to decrease by
increasing $z$. In general, the slope of the contours  in the
($c,\alpha$)-plane for the different $z$-bins is estimated by
calculating the variation of $1-R_M^h=c\langle A \rangle^{\alpha}$ at
an average  mass number  $\langle A \rangle$:  
\begin{align}
 \Delta(1-R^h_M)=\langle A \rangle^{\alpha}\Delta c+c\,\alpha \, \langle
 A\rangle^{\alpha-1} 
 \Delta \alpha.
\end{align}
A change $\Delta c$ of $c$ is correlated with a corresponding 
change $\Delta \alpha$ of $\alpha$ given by
\begin{align}
 \Delta \alpha = -\frac{\langle A \rangle}{c\,\alpha } \Delta c \ .
\end{align}
This equation gives the negative slopes of the contours. 
As the product of $c$ and $\alpha$ generally increases with
increasing $z$, this implies a decreasing slope.  

The best-fit parameters, with their uncertainties are given 
in Table~\ref{table:fitcoeff}, both for the fits of the experimental data and  the pure absorption model predictions. By showing the best-fit parameters of the pure absorption model computation we can quantitatively compare deviations from the leading order $A^{2/3}$ dependence predicted in section 3. 
The uncertainties of the best-fit parameters are determined by
projecting the contour plots on the $c$- and $\alpha$-axes
respectively. Therefore, here as well, the parameter uncertainties
correspond to a joined $90\%$ confidence interval in the case of
normally distributed parameters. 
      
\begin{table}[t]
\begin{center}
\begin{tabular}{|c|c|c|c|c|c|c|} \hline
    &  \multicolumn{2}{c|}{Experiment} & \multicolumn{2}{c|}{Theory}  & \multicolumn{2}{c|}{Theory} \\
    &  \multicolumn{2}{c|}{He (N) Ne Kr} & \multicolumn{2}{c|}{He (N) Ne Kr}  & \multicolumn{2}{c|}{He (N) Ne Kr Xe} 
    \\ \cline{2-7}
    $z$ & $c\;[10^{-2}]$ & $\alpha$  & $c\;[10^{-2}]$     & $\alpha$    & $c\;[10^{-2}]$ & $\alpha$
\\ \hline \hline
$.25$ &  $2.1\pme{0.8}{0.5}$ & $0.51\pme{0.06}{0.10}$ & $0.7\pme{0.9}{0.5}$& $0.75\pme{0.22}{0.20}$ & $0.9\pme{0.9}{0.4}$& $0.70\pme{0.15}{0.17}$ \\ \hline

$.35$ &  $2.6\pme{0.8}{0.6}$ & $0.47\pme{0.08}{0.07}$ & $0.7\pme{0.9}{0.4} $ & $0.77\pme{0.23}{0.20}$ & $0.8\pme{0.9}{0.4} $ & $0.72\pme{0.15}{0.17}$ \\ \hline

$.45$ &  $1.9\pme{0.7}{0.4}$ & $0.58\pme{0.09}{0.10}$ & $0.7\pme{0.8}{0.4}$ & $0.78\pme{0.22}{0.20}$ & $0.8\pme{0.9}{0.4}$ & $0.73\pme{0.15}{0.17}$ \\ \hline

$.55$ &  $1.6\pme{0.7}{0.6}$ & $0.62\pme{0.13}{0.10}$ & $0.8\pme{0.7}{0.4}$ & $0.76\pme{0.16}{0.17}$ & $0.9\pme{0.7}{0.4}$ & $0.71\pme{0.11}{0.13}$  \\ \hline

$.65$ &  $1.8\pme{1.0}{0.7}$ & $0.61\pme{0.13}{0.14}$  & $1.0\pme{0.8}{0.4}$ & $0.74\pme{0.14}{0.16}$ & $1.1\pme{0.8}{0.4}$ & $0.70\pme{0.10}{0.13}$ \\ \hline

$.75$ &  $1.3\pme{0.8}{0.6}$ & $0.72\pme{0.15}{0.13}$  & $1.2\pme{0.9}{0.5}$ & $0.73\pme{0.11}{0.15}$ & $1.4\pme{0.9}{0.4}$ & $0.68\pme{0.08}{0.13}$ \\ \hline

$.85$ &  $1.2\pme{0.5}{0.7}$ & $0.78\pme{0.22}{0.10}$ & $1.7\pme{1.2}{0.5}$ & $0.69\pme{0.09}{0.15}$ & $1.9\pme{1.2}{0.5}$ & $0.65\pme{0.06}{0.12}$ \\ \hline

$.95$ &  $3.6\pme{2.1}{1.3}$ & $0.56\pme{0.12}{0.12}$  & $3.1\pme{1.5}{0.8}$ & $0.60\pme{0.07}{0.12}$ & $3.3\pme{1.6}{0.7}$ & $0.57\pme{0.05}{0.10}$ \\ \hline
\end{tabular}
\vskip.3cm
\end{center}
\caption{Centroids of the contour plots  in
  Figs.~\ref{fig:fig6} and \ref{fig:fig7} with their uncertainties for
  the fit $1-R_M=c(\nu,z,h) A^\alpha$ at fixed $z$ bins,
  both for the experimental data and the pure absorption model
  calculation. The nuclei included in the fits are shown in the table,
  the parenthesis on (N) indicate that nitrogen is included in the fit
  only for  $z\ge0.55$. Note, that the theoretical results include different sets of nuclei.}    
\label{table:fitcoeff}
\vskip.3cm
\end{table}

\begin{figure}[ht]
\centering
\hspace*{-2.25cm}
\includegraphics[width=18.0cm, height=19.cm]{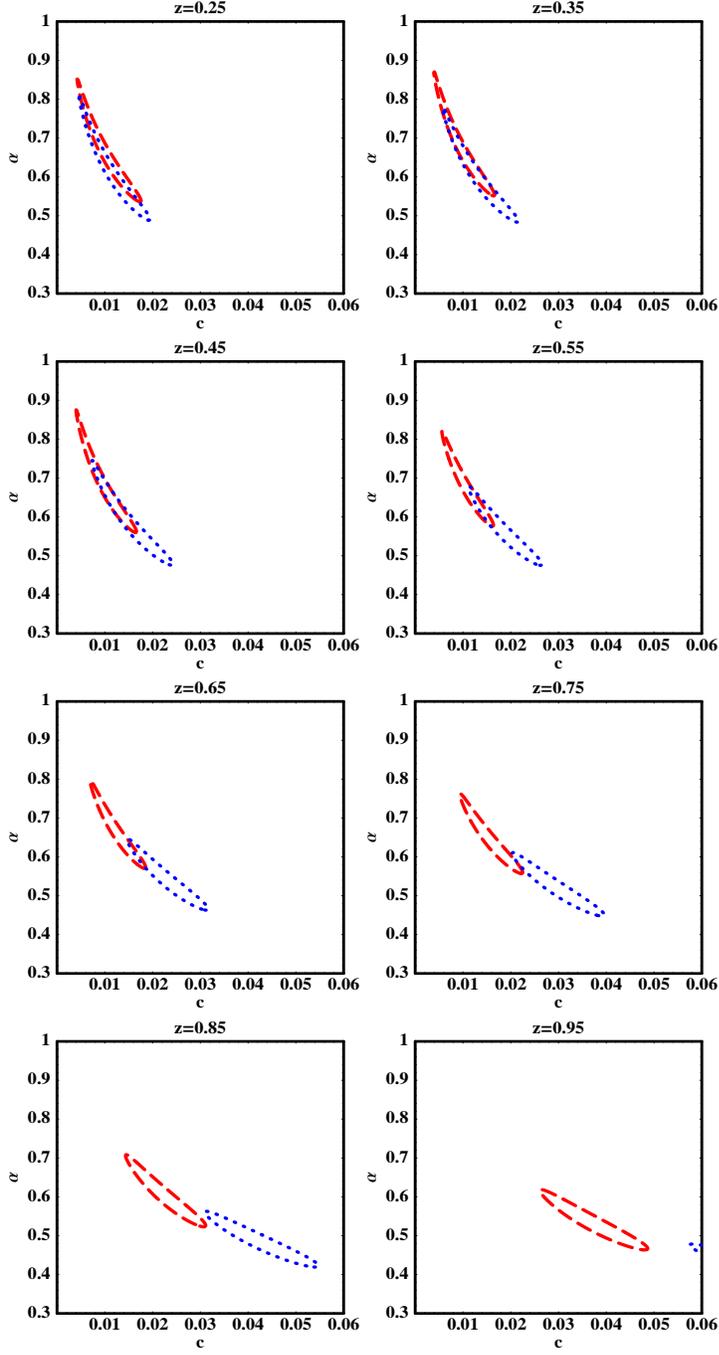}
\vskip-0.5cm
\caption{$\chi^2$ contour plots for the fit $1-R_M=c(\nu,z,h)
  A^\alpha$ on the pure absorption model (dashed) and the full model (dotted) computation for $^{4}$He, ($^{14}$N),
  $^{20}$Ne,  $^{84}$Kr and  $^{131}$Xe nuclei, in fixed $z$-bins. Note that the contour 
  for the full theory computation in the last $z$ bin is outside of
  the plot range, to the right. }
\label{fig:fig7}
\end{figure}

The values of $\alpha$ derived from both the pure absorption model fits 
and the experimental fits on He,($^{14}$N), Ne, Kr at $z\ge 0.55$ are
compatible, within their uncertainties, with  $A^{2/3}$
behavior. Deviation from this behavior should be observed for heavier
nuclei than krypton and at high $z$ as shown in
Table~\ref{table:c1c2}. In order to observe the breaking of the power
law predicted by the model and to increase the statistical
significance of the lowest $z$-bins, we perform the same computation
by including xenon nucleus. Experimental data on Xe have been recently
collected by the HERMES experiment, but the analysis is as yet in progress. The theoretical contour plots are shown in Fig.~\ref{fig:fig7} and the best-fit
parameters for the pure absorption model with their uncertainties are summarized in
Table~\ref{table:fitcoeff}. Since we include a nucleus with a mass
number comparable to $\bar{A}$ for large $z$ bins, as shown in
Table~\ref{table:c1c2}, the power law $A^{2/3}$ is broken
significantly producing a reduction of the $\alpha$ values. 

\begin{figure}[tbh]
 \centering
 \includegraphics[width=1.0\linewidth]{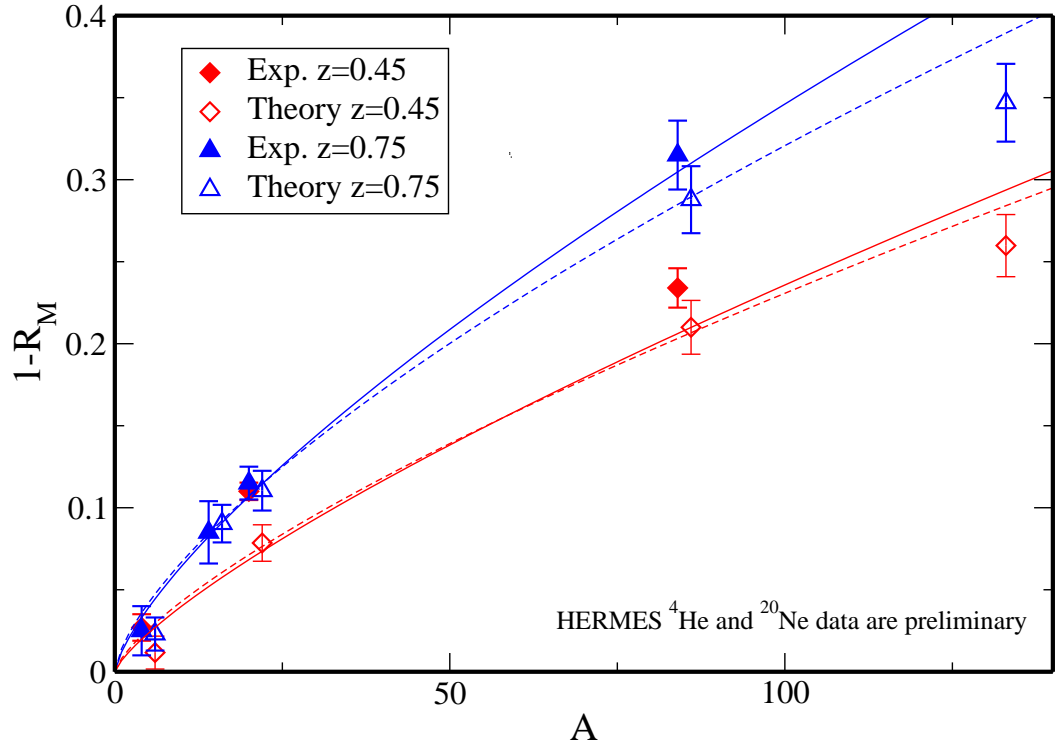}
 \caption{The published HERMES data for $R_M$ for $^{14}$N~\cite{HERM01} 
 and  $^{84}$Kr~\cite{HERM03} and the preliminary data for $R_M$ for 
 $^{4}$He~\cite{HERM04} and $^{20}$Ne~\cite{HERM04} are shown as $1-R_M$ for  
 $z=0.45$ and $z=0.75$ as filled diamonds and filled triangles respectively. 
 The pure absorption model results for $1-R_M$ in the same $z$-bins and
for the same nuclei plus $^{131}Xe$ are shown by empty symbols.
Note that we have shifted the absorption model values slightly to
the right to avoid overlap with the experimental points.  
 The solid lines are a best fit of the pure 
 absorption model results to $1-R_M=c A^\alpha$ including values for $A=4,14,20,84$ (He,N,Ne,Kr) and the dashed lines represent a fit including values for $A=4,14,20,84,131$ (He,N,Ne,Kr,Xe).  
}
 \label{fig:AttenVsA}
\end{figure}

In Fig.~\ref{fig:AttenVsA} the 
published HERMES data for $R_M$ for $^{14}$N~\cite{HERM01} and
$^{84}$Kr~\cite{HERM03} and the preliminary data for $R_M$ for
$^{4}$He~\cite{HERM04} and $^{20}$Ne~\cite{HERM04} are displayed as
$1-R_M$ for two selected $z$-bins ($z=0.45,0.75$) with full diamonds and
triangles. The empty symbols show the corresponding results of the pure 
absorption model for the above nuclei plus $^{131}Xe$.
In addition we plot with solid (dashed) lines 
the best fits to $1-R_M=c A^\alpha$ of the theory results including
He,N,Ne,Kr (He,N,Ne,Kr,Xe). 
One observes that the inclusion of xenon flattens the curve for the 
attenuation $1-R_M$ for large values of $z$. 
The exponent extracted by the proposed $(c,\alpha)$ fit can be
regarded as an average exponent on the considered interval of atomic
masses. Beyond $A\approx80$ and for large values of $z$ the xenon point
lies below the $A^\alpha$ curve fitted to all nuclei.

\begin{figure}[h]
\centering
\vskip-1cm
\includegraphics[width=1.0\linewidth]{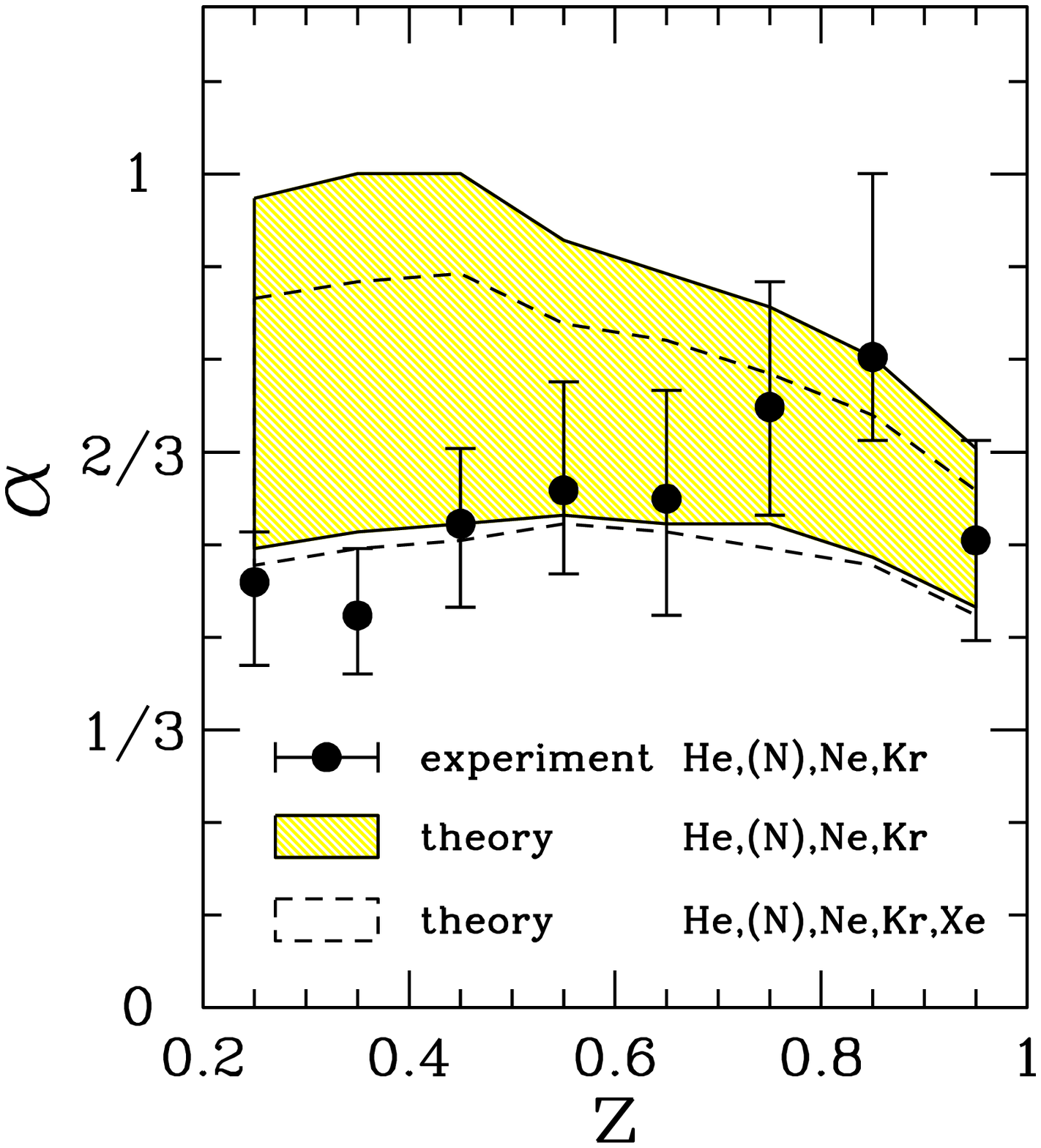}
\caption{Values of $\alpha$ as a function of $z$ derived from
  experimental data (dots) and our pure absorption model computation (bands).
  The nuclei included in the fit are shown  in the legend (the
  HERMES data on He and Ne are preliminary). The notation (N)
 indicates that this nucleus is included in the fit only
  at $z\geq 0.55$. Note that experimental errors are uncorrelated, but
  theory errors are point-to-point
  correlated.
 }
\label{fig:fig9}
\end{figure}

In order to investigate a breaking of the $A^{2/3}$ power law at large $A$ 
we show \figref{fig:fig9}, where the 
$z$-dependence of theoretical $\alpha$ values in our pure absorption
model is compared with the $z$-dependence of $\alpha$ values derived
from experimental data. 
The xenon nucleus shifts the upper band limit of the theoretical
predictions to lower $\alpha$ values. But the $\alpha$ values with Xe  
are compatible with the $\alpha$ values without the Xe nucleus over the whole $z$ range.
Nevertheless the $A$-dependence of hadron
attenuation in nuclei is a promising tool in order to
distinguish different theoretical assumptions.

\section{Summary and conclusions}
\label{sec:conclusions}

Rescaling of the structure and fragmentation functions together
with (pre) hadronic absorption describe both HERMES and EMC data on
the nuclear modification of hadron production in DIS processes
\cite{ACC02}. This paper improves the original model of \refref{ACC02},
by correctly evolving quarks, prehadrons and hadrons without
factorizing the absorption process and hadronization
processes. 
The hadron formation length is shown to be flavor
dependent. Especially, negative kaons ($K^-=s\bar{u}$) as well as
antiprotons ($\bar{p}=\bar{u}\bar{u}\bar{d}$) cannot be formed by a
struck valence quark which is the dominant contribution at HERMES
energy.  
This yields a smaller formation length for negative kaons and antiprotons 
implying more hadron suppression compared to positive kaons or
protons respectively.  
Furthermore the so far neglected transverse extension of the produced
hadrons leads to slightly different formation lengths for different
hadron species.
We now use a reasonable value for the  string tension $\kappa=1$ 
GeV/fm, and extract the prehadronic cross-section 
$\sigma_*=2/3\sigma_h$ from data on pion production on krypton.
This shows that the prehadron does not have the full hadronic
size yet. 

The presented model correctly describes pion and kaon multiplicity
ratios. A different magnitude of the multiplicity
ratios for positive and negative kaons originating from a flavor dependent
formation length is reflected in the experimental data. 
In the baryon sector, although computed antiproton mutliplicities agree with experimental data inside their uncertainties, model computations disagree with proton data especially at small $z$.
This discrepancy may be ascribed to a
non negligible diquark fragmentation contribution in that region.
Prehadron rescatterings and final state interactions are insufficient
to fully account for the effect. 

We have applied our model to the analysis of the mass number
dependence of the hadron attenuation $1-R_M$. Contrary to the common
expectation we found a $cA^{2/3}$ behavior in leading order,
with higher order corrections increasing in magnitude with increasing
$z$ and $A$.
Hence, one must not expect a strict $cA^{2/3}$ power law,  
but rather an effective behavior $cA^{\alpha}$. Furthermore, the
proportionality coefficient $c$ depends strongly on the kinematic
variables, which rules out a simple analysis
  with a globally fixed $c$. Therefore we  
have proposed  to analyze the
$A$-dependence of hadron attenuation in terms of a power-law fit 
\begin{align}
 1-R_M = c\; A^{\alpha}
\end{align}
with both $c$ and $\alpha$ as fit parameters. Their correlations,
displayed as $\chi^2$ contours in the $(c,\alpha)$-plane, are
quite sensitive to the underlying model assumptions and may
disentangle different theoretical models. 

Qualitative agreement is found between the theoretically
and experimentally determined $\chi^2$ contours, with
$\alpha\approx 2/3$. Contrary to common expectations, $\alpha\approx
2/3$ also in the presented absorption model. This value is not a
peculiarity of our model, but a common feature of a quite
large class of absorption models (a sufficient requirement being an
analytic probability distribution in the formation length $\Delta y$, 
with a finite limit as $\Delta y\ra 0$). This fact bars the much
advertised use of the exponent $\alpha = 2/3$ as a direct
experimental evidence of parton energy loss following the QCD
''$L^2$'' law. 

In different models the breaking of the $\alpha=2/3$ power law
happens at different values of $A$ and $z$. Therefore  
collecting data up to heavy
nuclei like Sn and Pb, which is in progress at Jefferson Laboratory
with the CLAS experiment \cite{Brooks}, 
will help unraveling the space-time dynamics of the
hadronization process. The proposed analysis of the $A$-dependence of
the hadron attenuation may thus provide a new method to
differentiate between absorption and energy loss effects. In this
respect we encourage the authors of competing models to perform a
similar $(c,\alpha)$ analysis to have a full set 
of theoretical results when the new data from HERMES and CLAS 
measurements will be available.

\vskip1cm

{\bf Acknowledgments} 

  We are grateful to P.~Di Nezza, C.~Ogilvie,
  J.~Raufeisen and M.~Rosati for fruitful discussions and many useful
  comments.
This work is partially funded by the US Department of Energy grant
DE-FG02-87ER40371 and by the european union project EU RII3-CT-2004-506078. 
\vskip.15cm

\vfill

\newpage
\begin{appendix}

\section{Generic absorption model}
\label{app:genericabs}

In this appendix we show that hadron attenuation  
behaves as $1-R_M = c_0 A^{2/3} +
\mathcal{O}(A^1)$ for a large class of absorption models 
and is not a peculiarity of our model. 
A sufficient requirement is that
the probability distribution of the prehadron formation length $\Delta
y$ is analytic and has a finite limit as $\Delta y \ra 0$.

We consider the generalized 1 step approximation described in
Sect.~\ref{sec:DevModel} by substituting in \eqeqref{1-rm} the
exponential probability distribution for prehadron formation with a
generic distribution $\PP$:
\begin{align}
  \frac{1}{\vev{l^*}} e^{-\frac{x-y}{\vev{l^*}}} 
    \ \lora \ \frac{1}{\Lambda} \PP\left(\frac{x-y}{\Lambda}\right) \ ,
\end{align}
where $\Lambda$ is a typical scale of the fragmentation process, and
$\PP(u)$ is normalized to 1:
\begin{align}
  \int_0^\infty \hspace*{-.2cm}du\, \PP(u) = 1 \ .
 \label{normaliz}
\end{align}
The absorption factor \eqref{BCsimple} is then generalized as follows:
\begin{align}
  1-R_M=\frac{\pi \rho_0}{2 A} \Lambda^3
    \int\limits_0^{2R/\Lambda} dt\,t
    \int\limits_0^t dr \int\limits_0^r du\, \PP(u)
    \left[ 1-e^{\frac{\Lambda}{\lambda_*} (u-r)} \right] 
 \label{genabs}
\end{align}
(note that integration over $t$ and $u$ is the integration
over all possible $\gamma^*q$ interaction points).
This generalization captures the essential features of the
$A$-dependence of the hadron suppression factor $1-R_M$ in most
absorption models found in literature. A few examples will be
discussed below. 

A quite general probability distribution can be defined as follows.
Let's $\PP$ be an analytic function on $(0,M)$ with $M>0$ or
$M=\infty$. Let's $n$ be the smallest real number such that
$\lim_{\,u\ra0} u^{-n} \PP(u) = p_0$ with $p_0\neq0,\infty$. Then the function 
\begin{align}
  f(u) = \left\{\begin{array}{ll}
    u^{-n} \PP(u) \ & u \in (0,M) \\
    p_0 & u=0
  \end{array} \right.
\end{align}
can be expanded in powers of $u$ around $u=0$:
\begin{align}
  f(u) = p_0 + p_1 u + \OO(u^{2})\ .
\end{align}
This means that
\begin{align}
  \PP(u) = p_0 u^n + \OO(u^{n+1})\ .
\end{align}
By the normalizability requirement \eqref{normaliz} the exponent
$n$ must satisfy
\begin{align}
  n>-1 \ .
 \label{nlargerminusone}
\end{align}
However, to our knowledge, there is no probability distributions
considered in literature with $n < 0$.
Now, if we expand the integrand of \eqref{genabs} in a Laurent series
around $u=0$ and perform all the integrations, we obtain
\begin{align}
  1-R_M = \frac{\pi \rho_0}{2 A}
    \frac{p_0 \, 2^{n+5}}{(n+1)(n+2)(n+3)(n+5)} 
    \frac{\Lambda^4}{\lambda_*} \frac{R^{n+5}}{\Lambda^{n+5}} 
    + \mathcal{O}\left(\frac{R^{n+6}}{\Lambda^{n+6}}\right)\ .
 \label{result}
\end{align}
If we now use $R=r_0 A^{1/3}$ we obtain
\begin{align}
  1-R_M = c_0 A^{\frac{n}{3}+\frac{2}{3}} +
  \mathcal{O}(A^{\frac{n}{3}+1}) \ ,
 \label{genexp}
\end{align}
where the constant $c_0$ is easily read off \eqeqref{result}.
We see that the exponent of the first-order term is always larger than
1/3 by virtue of \eqeqref{nlargerminusone}. However, since no models
in literature have $n<0$, the exponent is in practice always larger
than or equal to 2/3. 
From \eqeqref{genexp} we see that in absorption models such that
$\lim_{u\ra0} \PP(u) = p_0 > 0$, which have $n=0$, 
the expansion of $1-R_M$ in powers of $A$ starts with an $A^{2/3}$
term, as claimed. The exponential distribution considered in this paper
falls in this class of models. 

An example of a model with $n>0$ is the bremsstrahlung model of
\refref{KOP03}. In that paper the computed probability distribution
satisfies $\PP(u)\ra0$ as $u\ra0$, so that $n>0$ and  
\begin{align}
  (1-R_M)_{\cite{KOP03}} = c_{\cite{KOP03}} 
    A^{\alpha_{\cite{KOP03}}} + \mathcal{O}(A^{\alpha_{\cite{KOP03}}+\frac{1}{3}}) 
\end{align}
with $\alpha_{\cite{KOP03}}>2/3$. However, except for very small $u$
or small $z$, the probability distribution is well approximated by an
exponential distribution. 
Hence, for realistic finite-size nuclei we expect higher order
corrections to be important and to make the $A$ dependence of $1-R_M$
effectively close to $A^{2/3}$.  

Models with non-analytic probability distribution can have
different behaviour. For example, in \refref{FALTER04} all prehadrons
are assumed to be formed immediately after the interaction of the
virtual photon 
with a quark (though prehadrons created from the middle of the color
string are assumed not to interact with the medium). Therefore,
absorption of rank 1 prehadrons can be modeled as in \eqeqref{genabs}
with  
\begin{align}
  \PP(u) = \delta(u) \ .
\end{align}
Carrying out all the integrations in \eqref{genabs} one finds
\begin{align}
  (1-R_M)_{\cite{FALTER04}} = c_{\cite{FALTER04}} A^{1/3} 
\end{align}
with no higher order corrections. However, this $A^{1/3}$ 
power law may change when also considering the other processes  
included in the model, like diffractive meson production (important at
large $z$) and higher-rank prehadron production, elastic (pre)hadron
rescatterings, resonance decays (important at smaller $z$).

On the other hand, a non-analytic probability distribution is not a
guarantee for an $A^{1/3}$  power-law. Indeed, the Bialas-Chmaj model
(BC) of \refref{BICH83} can be shown \cite{ACC02} to reduce to
\eqeqref{genabs} with $\Lambda = \vev{l^*}$ and
\begin{align}
  \PP(u) = \delta[u - (1-e^{-r})] \ . 
\end{align}
Carrying out all the integrations in \eqref{genabs} one finds
\begin{align}
  (1-R_M)_{\cite{BICH83}} = c_{\cite{BICH83}} A^{2/3} 
    + o(A^{2/3}) \ . 
\end{align}

This concludes our proof that an $A^{2/3}$ power law for the hadron
suppression factor of not too heavy nuclei is a general
feature of a quite large class of absorption models.  

\end{appendix}


\bibliographystyle{unsrt}

\end{document}